\documentclass[journal]{IEEEtran}

\usepackage{amsmath} 
\usepackage{amsfonts} 
\usepackage{amssymb}
\usepackage{amsthm}
\usepackage[utf8]{inputenc}
\usepackage{graphicx}
 \usepackage[caption=false,font=footnotesize]{subfig}
\usepackage{bbm}
\usepackage{csquotes}
\usepackage{enumerate}
\usepackage[numbers,sort&compress]{natbib} 
\usepackage{color}
\usepackage[normalem]{ulem}
\usepackage{numprint}
\usepackage{fixltx2e}
\usepackage{url}
\usepackage{hyperref}

\newtheorem{cor}{Corollary}
\newtheorem{theo}{Theorem}
\newtheorem{rem}{Remark}
\newtheorem{lem}{Lemma}

\IEEEoverridecommandlockouts

\begin{document}

%\IEEEauthorrefmark{1}

\title{Ranking Online Social Users by their Influence}

\author{Anastasios~Giovanidis, %,~\IEEEmembership{Member,~IEEE,} 
	Bruno~Baynat, 
	Cl\'emence~Magnien,
	and~Antoine~Vendeville
\thanks{The preliminary version of this article appeared in the Proc. of INFOCOM 2019 \cite{INFOCOM19}. This work is funded by the ANR (French National Agency of Research) partly by the ``FairEngine'' project under grant ANR-19-CE25-0011, and partly by the ``Limass'' project under grant ANR-19-CE23-0010.
The authors are with the Sorbonne Universit\'e, CNRS, LIP6, F-75005 Paris, France, Email: \{firstname.lastname\}@lip6.fr
}
%\thanks{Manuscript received ...}
}

%\markboth{submitted to IEEE/ACM Transactions on Networking}%
%{Giovanidis \MakeLowercase{\textit{et al.}}}

\maketitle

\begin{abstract}
We introduce an original mathematical model to analyse the diffusion of posts within a generic online social platform. The main novelty is that each user is not simply considered as a node on the social graph, but is further equipped with his/her own Wall and Newsfeed, and has his/her own individual self-posting and re-posting activity. As a main result using our developed model, we derive in closed form the probabilities that posts originating from a given user are found on the Wall and Newsfeed of any other. These are the solution of a linear system of equations, which can be resolved iteratively. In fact, our model is very flexible with respect to the modelling assumptions. Using the probabilities derived from the solution, we define a new measure of per-user influence over the entire network, the $\Psi$-score, which combines the user position on the graph with user (re-)posting activity. In the homogeneous case where all users have the same activity rates, it is shown that a variant of the $\Psi$-score is equal to PageRank. Furthermore, we compare the new model and its $\Psi$-score against the empirical influence measured from very large data traces (Twitter, Weibo). The results illustrate that these new tools can accurately rank influencers with asymmetric (re-)posting activity for such real world applications.
\end{abstract}


\IEEEpeerreviewmaketitle

\section{Introduction}

\IEEEPARstart{O}nline Social Platforms (OSPs) play a major role in the way individuals communicate with each other, share news and  get informed. Today such platforms host billions of user profiles. Although OSPs differ from one another, most of them share a common structure, which allows users to post messages on their Wall and read posts of others on a separate Newsfeed. Most OSPs also permit re-posting from Newsfeed to Wall, in order to facilitate information diffusion. With each re-post (or ``share'', or ``re-tweet'') the information becomes visible to a new audience, which may choose to adopt it or not, thus spreading further the post or halting its diffusion. In this way, posts originally generated by some user circulate inside the social network \cite{EC2012}. When the post is gradually adopted by a considerable proportion of the users, we see large cascades of information appear, and we call such posts ``viral'' \cite{Adamic2013}.

Understanding how information spreads through OSPs is very important as it affects the opinion of the population over several subjects of every-day social life. Companies want to determine the set of most influential users (``influencers") for better marketing of their products \cite{Kempe03}, and they would like to predict information cascades \cite{LescovecPredict}. Such research is critical also because spreading of influence can have malevolent purposes instead \cite{FakeNewsPaper18}, such as the spread of misinformation (``fake news''). To be able to develop defence mechanisms against such social attacks, a concrete mathematical analysis of post diffusion through OSPs is necessary.

Existing literature on the topic has mainly focused on models for opinion dynamics that take only the social graph as input. These include the \textit{voter} model \cite{HoLi75}, the SI(R) \cite[Ch. 17]{NetworksBook},  the \textit{threshold and cascade} models \cite{Kempe03}, bootstrap percolation \cite{Leonardi16}, and the \textit{DeGroot} \cite{DeGroot} model, among others. In each of these, the social graph structure together with simplified user interaction, has been assumed sufficient to describe the diffusion \textit{of a single opinion}. However,  the authors of the highly cited paper about the ``million follower fallacy" \cite{1MfollowerF}, argue that graph topological measures alone reveal very little about the true influence of a user in a platform; they use large traces from Twitter to support their claim. The authors in \cite{SalaInteractionGraph} further use Facebook data to identify real indicators of user interactions, beyond social links. Our paper has the ambition to fill this gap between analysis and data-driven conclusions by introducing a new dynamic model which combines the information over the social graph together with user activity and the OSP structure, in order to explain more accurately how posts from different origins diffuse and compete among each other inside the social platform.

Viral marketing wants to identify users with high social influence \cite{Kempe03}. To this aim, users are ranked based on certain impact measures, which mainly depend on user graph position (e.g. number of follower links), similar to the existing opinion models discussed above. For example, \cite[Ch. 7]{NetworksBook} presents degree, eigenvector and Katz centrality, as well as PageRank score \cite{pagerank98}, and \cite{papadim11} alternatives for large-scale graphs. We claim that such measures are not suitable to rank the influence of social users, because they do not include user activity, or OSP structure and they will hence mislead when used to identify ``influencers". We propose instead a new $\Psi$-score to rank users, based on our proposed model. 

\subsection{Main Contributions and Paper Structure}
Our main contributions are summarised as follows:

\begin{itemize}
\item The entire OSP is described as a continuous-time Markov chain. This model is original in the sense that it combines (i) the social graph, with (ii) dynamic user posting and re-posting activity, and incorporates elements of (iii) the platform structure (Walls, Newsfeeds and the Newsfeed suggestion algorithm). The model can include various Newsfeed mechanisms (First-In-First-Out (FIFO), Random, Time-To-Live (TTL)) and user post sharing behaviour. Also, competition among posts to gain the user's attention on the Newsfeed is naturally included.

\item By analysing the above chain we result in a linear system of equations, which exactly describes the chain's behaviour in steady-state (Theorem~\ref{exact}, Theorem~\ref{th1}). This has as unknown variables the influence of a given user on the Wall and Newsfeed of any other. This system actually consists of the balance equations of posting activity on each Wall and Newsfeed and can be solved for an arbitrary input graph and arbitrary user activity rates. Theorem~\ref{Main} provides its solution.

\item An iterative method (Theorem~\ref{th3}) proposed to compute the system solution facilitates numerical implementation. It allows to implement a sparse algorithm, which scales well as the size of the social graph increases.

\item The solution gives rise to a new way to rank OSP users by their influence. We call the new ranking metric, the ``$\Psi$-score''. The performance of our model and the $\Psi$-score is tested on two large real-world traces from famous social platforms, one from Twitter and another from Weibo.

\item We prove in Theorem~\ref{pagerankQ} that in the homogeneous activity case where all users post and repost with the same rate a variant of the $\Psi$-score coincides with PageRank.
\end{itemize}
The implementation code is made available online \cite{code}.

The paper is organised as follows. The social platform under study and the performance metrics of interest are introduced in Section \ref{systemd}. The Markovian model describing the generic OSP and its balance equations are given in Section \ref{models}. Here, we explain how these linear equations naturally constitute the Newsfeed and Wall balance equations, and show their exactness and general validity. The system's closed-form solution is provided in Section \ref{closed}. In the same section we provide an iterative method that is computationally cheap and converges to this solution. In Section \ref{PageRank} we discuss relations with PageRank. The ranking algorithm based on the $\Psi$-score and its implementation for large data-traces is detailed in Section \ref{algo1}. Extended numerical experiments using synthetic data in Section \ref{numanal} verify the model's validity and robustness over modelling assumptions. Massive real-world traces from Twitter and Weibo are used to evaluate the $\Psi$-ranking and the sparse algorithm, in Section \ref{trace}. Our $\Psi$-ranking is further compared for these traces against standard user ranking metrics (number of followers, PageRank, and user posting rate). Conclusions are drawn in Section \ref{conclu}. The proof of exactness (Theorem~\ref{exact}) is made available in the Appendix.

\subsection{Related Literature}

In most relevant research on opinion dynamics, individuals are seen as agents whose relation is described by a social graph. Each agent has a certain opinion and at each step this opinion is updated through interaction with his/her direct peers. Such models can be grouped into two general categories.

\textit{1) Dynamics with Binary opinions:} There are only two possible opinions that agents can take.  A large amount of work descends from the \textit{voter model} \cite{HoLi75}, where opinion dynamics are based on imitation. The work in \cite{YOASS} studies a variation that includes agents with persisting opinions. For further extensions, see also \cite{HayelINFO18}, \cite{Mazum16}. Another group of work is related to epidemic spread. An agent is ``susceptible'' when his/her opinion is $0$ and becomes ``infected'' when he/she adopts opinion $1$, through social interaction \cite{NetworksBook}. In \cite{Kempe03} two opinion update mechanisms are studied: the threshold and the cascade.

\textit{2) Dynamics with Continuous opinions:} Several works in the literature have inherited and extended the original model of DeGroot \cite{DeGroot}. In this, each agent updates his/her continuous opinion by forming per-step a weighted linear combination of the current opinions of his/her peers. Variations of this model consider the inclusion of persistent agents \cite{Emily18}. In \cite{Silva17} this update mechanism is used to formulate and solve an opinion manipulation problem. To account for more realistic social behaviour, the authors in \cite{BaccINFO15} consider opinion dynamics where agents interact in pairs only when their opinions are already close.

\textit{Data, OSPs, and Cascades:} Instead of modelling opinion dynamics, recent works rather use available data to investigate more practically how posts spread within OSPs. The authors in \cite{EC2012} describe diffusion patterns that arise in specific online domains. Data analysis of large Facebook cascades is performed in \cite{Adamic2013}. Interestingly, the authors in \cite{LescovecPredict} propose ways to predict cascade growth using machine learning tools. The insufficiency of using graph-based only information to evaluate user influence is studied in \cite{1MfollowerF} and \cite{SalaInteractionGraph}.

\textit{User activity:}  In \cite{WhenP} the authors identify user activity as an important control tool for influence maximisation. Making extensive use of datasets, they study the appropriate times for a user to post or re-post in an OSP in order to maximise the probability of audience response. An interesting analytical effort to relate user activity with OSP design and post diffusion is made in \cite{SmartB16}. The authors use temporal point processes to model posting and re-posting activity of a user. They highlight the importance of the Newsfeed in post propagation and map user activity to post visibility, building on the idea that a post can be adopted by a follower when it is visible on his/her Newsfeed and not pushed away by competing posts. Their model, however, treats only a single user Newsfeed and does not consider the dynamics of the entire social graph. Furthermore, the dynamics of the Newsfeed list are inaccurately mimicked by a FIFO queue. Another relevant line of research includes \cite{TimelinesMenache} and \cite{Har19}, where the authors study the bias of Facebook's News Feed algorithm. They consider a bipartite graph of a set of users following a set of publishers, and model post activity as Poisson. Newsfeeds are here again approximated by infinite queues with TTL or FIFO service.

Compared to these works, we propose here a more correct and complete OSP model; we accurately model Newsfeeds as lists, we consider here an arbitrary graph of any size and include re-posting -- among other realistic features. Finally, we verify our model's validity by large real-world data traces.

\section{System description}
\label{systemd}

\begin{figure}[t!]
\centering
\includegraphics[width=0.8\linewidth]{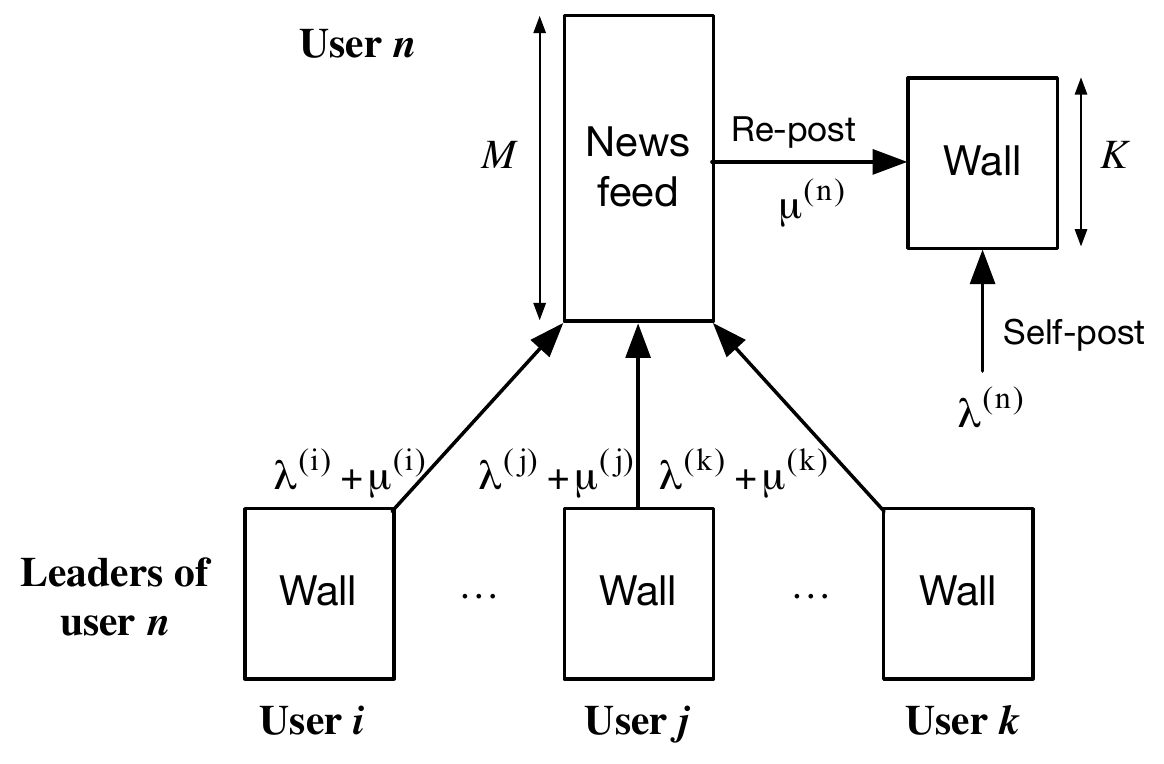}
\caption{The social platform from the point of view of user $n$.\label{system}}
\end{figure}

Let us first describe a generic social network platform, such as Facebook, Twitter or Weibo. A set of users generate and share some content, denoted as \textit{posts}, through the platform. Each user has a list of \textit{followers} and a list of \textit{leaders}. A user can simultaneously be follower and/or leader of others. As a follower, he/she is interested in the content posted by his/her leaders. With each user two lists of posts are associated, namely a \textit{Newsfeed} and a \textit{Wall}.
A user's Newsfeed is constantly fed by the content that all of his/her leaders post on their Walls. A user's Wall is fed (i) by his/her self-generated posts that draw influence from the ``outside world'', and (ii) by posts that he/she shares from his/her Newsfeed. Hence, a user's Wall is a list of self-posts and re-posts. The generic social network platform is illustrated in Figure~\ref{system}.

\subsection{Assumptions on the system and notations}

We consider a constant number $N$ of active users, forming the set $\mathcal{N}$. Users are labelled by an index $n=1,\ldots,N$. We denote by $\mathcal{F}^{(n)}$ and $\mathcal{L}^{(n)}$ the list of followers and the list of leaders of user $n$. Without loss of generality, we draw the directed Follower-graph $\mathcal{G}=(\mathcal{N},\mathcal{E})$. Each pair of nodes $(j,i)\in\mathcal{E}$, corresponds to a directed edge from $j$ to $i$, when $j$ is a follower of $i$, i.e., $j\in \mathcal{F}^{(i)}$. Such graph points to leaders. We denote by $\mathbf{F}$ the $N\times N$ adjacency matrix of the Follower-graph, whose coefficients are given by: $f_{j,i}=\mathbf{1}_{\left\{j\in \mathcal{F}^{(i)}\right\}}$, where $\mathbf{1}_{\left\{.\right\}}$ is the indicator function. We assume that each user $n$ has at least one leader, $\mathcal{L}^{(n)} \neq \emptyset$, $\forall n$. The Leader-matrix is by definition $\mathbf{L}:=\mathbf{F}^T$, so that $\ell_{i,j}=f_{j,i}$. Note, that the case of self-loops is excluded, i.e. $\ell_{i,i}=f_{i,i}=0$, $\forall i\in\mathcal{N}$.

The sizes of both Wall and Newsfeed are considered to be constant. We thus fix $K\geq 1$ the size of a Wall (total number of posts on the Wall of each user) and $M\geq 1$ the size of a Newsfeed. This is reasonable if we assume that only a certain number of most recent posts is considered relevant, and users don't tend to scroll down to access older posting history.

We denote by $\lambda^{(n)}$ $\mathrm{[posts/unit~time]}$ the rate with which user $n$ generates new posts on his/her Wall, and by $\mu^{(n)}$ the rate with which user $n$ visits his/her Newsfeed and selects one of the $M$ entries to re-post on his/her Wall (note here that each visit implies re-posting). As a result, posts arrive on the $n$-th Wall with a total rate $\lambda^{(n)}+\mu^{(n)}$ $\mathrm{[posts/unit~time]}$. Additionally, we make the assumption that content posted on a users's Wall instantaneously appears on the Newsfeeds of his/her followers. As a result,  the input rate of posts in the $n$-th Newsfeed, is $\sum_{j \in \mathcal{L}^{(n)}} ( \lambda^{(j)}+\mu^{(j)})$. Given that the two lists associated per user have fixed size, then with each new entry one element has to be removed from the list and replaced by the new one.
For the user activity we require $\lambda^{(n)}+\mu^{(n)}>0$, $\forall n$. 

Finally, any post originally generated by a given user $n$ takes as label the author's index $n$, and will keep this label throughout its lifespan inside the network.

\subsection{Influence metric of interest}
\label{metrics}

The aim is to estimate the influence of a specific user, say user $i$, over the entire network. In order to define the metric of interest, we first define the \textit{influence} of user $i$ on user $n$, denoted by $q_i^{(n)}$, as the expected percentage of posts of origin $i$ found on the Wall of user $n$. They obviously satisfy for each Wall $n$, $\sum_{i=1}^{N}q_i^{(n)}=1$, $\forall n$. We can also interpret $q_i^{(n)}$ as the probability that, when picking at random a post from Wall $n$, this post is of origin $i$. These performance quantities will be the output of the developed models. With the above, we propose the following metric of influence,
\begin{eqnarray}
\label{aim1}
\Psi_i & = & \frac{1}{N-1} \sum_{n \neq i} q_i^{(n)} \ \ \  \in\left[0,1\right].
\end{eqnarray}
It corresponds to the average percentage of posts of origin $i$ on the Walls of any user $n\neq i$. The suggested metric averages over all users in the network, but excludes the original user $i$. It holds $\sum_{i=1}^N\Psi_i<N/(N-1)$. Although the defined measure is somewhat natural, we will see that a similar measure 
\begin{eqnarray}
\label{prQ}
\tilde{\Psi}_i = \frac{1}{N}\sum_{n=1}^Nq_i^{(n)} = (1-\frac{1}{N})\Psi_i + \frac{1}{N}q_i^{(i)},
\end{eqnarray} 
which does not exclude self-influence is very important, as it coincides with PageRank in the homogeneous activity case $\lambda^{(n)}=\lambda$ and $\mu^{(n)}=\mu$ for all $n$ (see Theorem~\ref{pagerankQ}). For this variation $ \sum_{i=1}^N\tilde{\Psi}_i = 1$. Other metric definitions are also possible. As an example, we could use the probability to find at least one post of label $i$ on the Wall of user $n$. This is equal to $1-(1-q_i^{(n)})^K$, where the independence among slots holds for specific user behaviour and post replacement policies (e.g. random selection/eviction) as we will see next. Based on this we can define an alternative metric. 
In any case, by associating an influence score to each user, the social users can be ranked by decreasing order of their influence. From now on, we will call the expression in (\ref{aim1}) (resp. (\ref{prQ})) which quantifies the influence of a user in the platform, the \textit{$\Psi$-score (resp. $\tilde{\Psi}$-score) of user $i$}. In this work we will focus only on this metric, leaving others for future investigations.

\section{Model}
\label{models}

\subsection{Markovian model}
\label{assumptions}

The model relies on the following assumptions:

\begin{itemize}

\item \textbf{Poisson arrivals}. For any user $n$ the generation of new posts on his/her Wall follows a Poisson process with rate $\lambda^{(n)}$ and the re-posting activity from his/her Newsfeed follows a Poisson process with rate $\mu^{(n)}$.

\item \textbf{Random selection}. When a user visits his/her own Newsfeed, we assume that he/she selects at random one of the $M$ entries to re-post on his/her Wall.

\item \textbf{Random eviction}. A novel entry on the Wall or Newsfeed list will push out an older entry of random position. 

\end{itemize}

Thanks to these assumptions, the resulting models developed in the following are Markovian. Indeed, all inter-arrival times between posts and re-posts are exponential and all choices are probabilistic. The \textit{random selection/random eviction} policy will be assumed throughout the solution process to derive the Newsfeed and Wall balance equations. \textit{Random selection} models the case where users pick a post at random from their Newsfeeds, i.e. without order of preference. We will show later that our solution is actually robust to other selection choices, like the \textit{newest selection} where a user always picks up the object from the top of his/her Newsfeed list (see Section~\ref{pol}). \textit{Random eviction} can model platforms which put new posts to Newsfeeds (and less realistically to Walls) in a random order. This could model the Facebook News Feed, where content curation algorithms decide the order of content appearance based on some background machine learning algorithms. In Twitter, however, both Newsfeeds and Walls normally show posts in a First-In-First-Out fashion, so the appropriate policy in this case would be the \textit{FIFO eviction} i.e., the oldest object is removed from the list and the fresh content is placed at the top. We will show in Section~\ref{pol} that this eviction policy (and others) satisfies the same balance equations as the random one. Alternative options for the selection and eviction policies will be compared to our solution by simulation in Section~\ref{robust}.  Generalisation of our solution to realistic user behaviour or Newsfeed mechanisms is an important topic for future research.

\subsection{Detailed model}
\label{State}
The full state-description for this system is an N-tuple $\mathbf{U}:=(\mathbf{U}^{(1)},\ldots,\mathbf{U}^{(N)})$, where $\mathbf{U}^{(n)}=(\mathbf{x}^{(n)},\mathbf{y}^{(n)})$ is the state of user $n$ (at a given time $t$, omitted in notations for sake of clarity). $\mathbf{x}^{(n)}$ is the state of his/her Newsfeed and $\mathbf{y}^{(n)}$ the state of his/her Wall. The random eviction and random selection assumptions allow to describe the system-state evolution without using information over the order of posts in the lists. Then, $\mathbf{x}^{(n)}=(x_1^{(n)},\ldots,x_N^{(n)})$, where $x^{(n)}_i$ counts the number of posts with user-origin $i$ found on the Newsfeed of user $n$.
Similarly,  $\mathbf{y}^{(n)}=(y_1^{(n)},\ldots,y_N^{(n)})$, where $y^{(n)}_i$ counts the number of posts with origin $i$ found on the Wall of user $n$.

With all the assumptions described in Section~\ref{assumptions}, the stochastic process with full state $\mathbf{U}$ is a continuous-time Markov chain model with finite state-space. This process obtains a unique stationary distribution. However, even for very small values of the system parameters the number of states will be enormous, whereas the state of a user's Newsfeed and Wall is coupled with the state of other users. As a result, any numerical method to find the solution, would be computationally intractable. For this reason we first introduce in the next subsections a state aggregation and a simple decoupling of the state-space that considerably reduce the solution complexity. Following that, we prove that the resulting balance equations we find are exact for the detailed model.

It is important to understand where the coupling between states of different users appears in the detailed model, before presenting the aggregated and decomposed model. Consider user $n$ and focus on label $i$ posts. A leader $k$ of user $n$ will re-post from his/her own Newsfeed to his/her own Wall a post of label $i$ with probability $x_i^{(k)}/M$, due to the random selection policy. This post will appear immediately on the Newsfeed of user $n$, thus changing its state $\mathbf{x}^{(n)}$. Hence, the evolution of the state of user $n$ depends not only on his/her own current state and on his/her own activity, but also on the current states of his/her leaders (in this example $x_i^{(k)}$).

Note here that the Markov process which describes the evolution of system state $\mathbf{U}$ is non-reversible \cite{Kelly}. To see this, we can use a simple example, given in Appendix A.

\subsection{State aggregation}

To simplify the solution process we first need to describe the state-space in a more compact way. To do so, we focus on posts from a particular user $i$ and calculate the influence of this user $i$ on the entire network. Of course, one can successively apply the technique to all $i=1,\ldots,N$ in order to determine eventually the influence and $\Psi$-scores of everyone.

The state aggregation is as follows. On all $N$ Walls and $N$ Newsfeeds we consider only two types of posts; those of origin $i$, and those issued from other users labelled as $-i$. In other words we aggregate the effect of all users except $i$. Remember that the detailed state of user's $n$ Newsfeed was the $N$-dimensional vector $\mathbf{x}^{(n)}$. By applying the state-aggregation this is now described by $(x_i^{(n)},x_{-i}^{(n)})$, whose sum is equal to the Newsfeed size $M$, so that $x_{-i}^{(n)}=M-x_i^{(n)}$. As a result, the state of the Newsfeed of user $n$ is reduced to a single integer $\mathbf{x}^{(n)}=x_i^{(n)}$ with values ranging from $0$ to $M$. Similarly, the state of user $n$'s Wall becomes also 1-dimensional $\mathbf{y}^{(n)}=y_i^{(n)}$ with values ranging from $0$ to $K$.

\subsection{Decomposition by mean-field approximation}

After state aggregation, the states $(x_i^{(n)},y_i^{(n)})$ of different users $n$ are always coupled among each other. We decompose here the state-description, to obtain $2N$ independent 1-dimensional
Markov Chains, each one associated with the Newsfeed and the Wall of a user. To do so, we use a ``mean-field'' approximation \cite{GastHdR20}: for a given user $n$, the state transitions of his/her Newsfeed and Wall will still be a function of his/her own current state and activity, as well as the activity of all of his/her leaders. But they will not depend anymore on the current Newsfeed and Wall states of the user's leaders $x_i^{(k)}(t)$, $y_i^{(k)}(t)$ but rather on their \textit{average probabilities in steady-state}, which at this stage are unknown values. 

More precisely, let us consider user $n$. We denote by $p_i^{(n)}$ the steady-state probability for a post on Newsfeed $n$ to be of label $i$, i.e., to originate from user $i$. Similarly, we have already defined in Section~\ref{metrics} $q_i^{(n)}$ as the steady-state probability for a post on the Wall of user $n$ to be of label $i$. These quantities for $n=1,\ldots,N$ are the model \textit{unknowns} after aggregation. Note that these probabilities are actually the ergodic means of the related user states, i.e., $p_i^{(n)} = \mathbb{E}[\frac{X_i^{(n)}}{M}]$ and $q_i^{(n)} = \mathbb{E}[\frac{Y_i^{(n)}}{K}]$.

We distinguish here between the Newsfeed and Wall of user $i$ (particularized user) and the Newsfeeds and Walls of users $n=j\neq i$. Consider the Newsfeed of user $j$; the state $x_i^{(j)}$ can evolve as follows:
\begin{eqnarray}
\label{Sxi}
x_i^{(j)}\stackrel{f_+^{(j)}(x_i^{(j)})}{\longrightarrow} x_i^{(j)}+1 \ \ \ & \& &\  \ \ x_i^{(j)}\stackrel{f_-^{(j)}(x_i^{(j)})}{\longrightarrow} x_i^{(j)}-1.\nonumber
\end{eqnarray}
In the above $f_+^{(j)}$ and $f_-^{(j)}$ are the transition rates between the states, respecting the range $0$ to $M$. The rate $f_+^{(j)}(x_i^{(j)})$ is
\begin{eqnarray}
\label{f+}
f_+^{(j)} = \left(\lambda^{(i)}\mathbf{1}_{\left\{i\in\mathcal{L}^{(j)}\right\}} + \sum_{k\in\mathcal{L}^{(j)}}\mu^{(k)}p_i^{(k)}\right)\frac{M-x_i^{(j)}}{M}.
\end{eqnarray}
Indeed, $\lambda^{(i)}$ is the rate with which user $i$ generates a new own post on his/her Wall, post that instantaneously appears on the Newsfeed of user $j$, if $i \in \mathcal{L}^{(j)}$.
A post of label $i$ can as well appear in the Newsfeed of $j$ through re-posting. This occurs when one of the $j$'s leaders visits his/her own Newsfeed and re-posts a label $i$ post on his/her own Wall. For leader $k$ such event occurs with rate $\mu^{(k)}\frac{x_i^{(k)}}{M}$, following the random selection policy. We propose that $\mu^{(k)}p_i^{(k)}$ is an estimation of this rate and this is where the \textit{``mean-field'' approximation} that decomposes the state-space lies. Finally, $\frac{M-x_i^{(j)}}{M}$ is the probability that an incoming post (with label $i$) replaces an old post of label $-i$, by the principle of random eviction.

\begin{figure*}
\centering
\includegraphics[width=\textwidth]{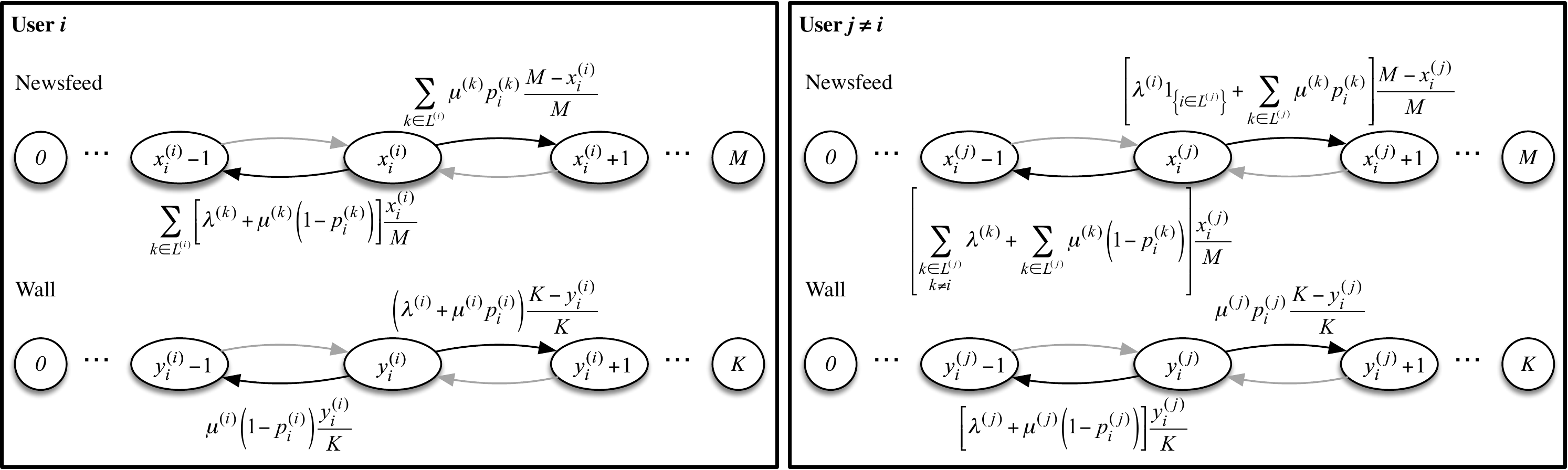}
\caption{Aggregated continuous-time Markov Chain model results in $2N$ birth-and-death processes.
\label{BDP}}
\end{figure*}

The rate $f_-^{(j)}(x_i^{(j)})$ is defined in a similar manner as
\begin{eqnarray}
\label{f-}
f_-^{(j)} = \left(\sum_{k\in\mathcal{L}^{(j)},k\neq i}\lambda^{(k)}+ \sum_{k\in\mathcal{L}^{(j)}}\mu^{(k)}(1-p_i^{(k)})\right)\frac{x_i^{(j)}}{M}.
\end{eqnarray}
As a consequence, the rate transitions from $x_i^{(j)}$ to $+1$ or $-1$ depend only on the current state of Newsfeed $j$ and not that of its leaders. We can thus describe $x_i^{(j)}$ as a 1-dimensional Markov Chain. Similar arguments hold for the Newsfeed state of user $n=i$. These simple Markov Chains are shown at the top part of Figure \ref{BDP}.
%%%==================== MARKOV CHAIN ====================%%%
%%%======================================================%%%
From the stationary probabilities of the chain for Newsfeed $n$, we can derive the steady-state probabilities $p_i^{(n)}$:
\begin{equation}
\label{p}
p_i^{(n)} = \sum_{x_i^{(n)}=0}^{M} \pi(x_i^{(n)}) \frac{x_i^{(n)}}{M}.
\end{equation}
Observe that the unknown probabilities $p_i^{(n)}$ depend on the steady-state solution of the 1-dimensional Markov chain, whose transition rates depend in their turn on the probabilities $p_i^{(.)}$ (see e.g. (\ref{f+})). As a result, the probabilities $p_i^{(n)}$ result from the solution of a \textit{fixed-point problem}.

For the Walls of users we can follow a similar process. In the Wall of user $j\neq i$, the state $y_i^{(j)}$ can evolve as follows:
\begin{eqnarray}
\label{Syi}
y_i^{(j)}\stackrel{g_+^{(j)}(y_i^{(j)})}{\longrightarrow} y_i^{(j)}+1 \ \ \ & \& &\  \ \ y_i^{(j)}\stackrel{g_-^{(j)}(y_i^{(j)})}{\longrightarrow} y_i^{(j)}-1.\nonumber
\end{eqnarray}
In the above $g_+^{(j)}$ and $g_-^{(j)}$ are the transition rates between the states, respecting the range $0$ to $K$. The rate $g_+^{(j)}(y_i^{(j)})$ is
\begin{eqnarray}
\label{g+}
g_+^{(j)}(y_i^{(j)}) & = &  \mu^{(j)}p_i^{(j)}\frac{K-y_i^{(j)}}{K}.
\end{eqnarray}
Indeed, the state of posts $i$ on the Wall of user $j\neq i$ can only evolve by reposting. Such posts enter the Wall $j$ with average rate $\mu^{(j)}p_i^{(j)}$, because the user re-posts with rate $\mu^{(j)}$ and has $p_i^{(j)}$ probability to choose posts of label $i$, due to random selection. This is again the \textit{``mean-field'' approximation}. The incoming post will replace an old post of label $-i$ with probability $\frac{K-y_i^{(j)}}{K}$ due to random eviction. The corresponding rate $g_-^{(j)}(y_i^{(j)})$ is defined in a similar fashion. Hence, the state evolution of posts $i$ on the Wall of user $j$ can be described by an independent 1-dimensional Markov Chain. We proceed similarly for the Wall of user $n=i$. These simple Markov chains are illustrated in detail at the bottom part of Figure~\ref{BDP}. From the stationary probabilities of the chain associated with user's $n$ Wall, we can derive the $q_i^{(n)}$:
\begin{equation}
\label{q}
q_i^{(n)} = \sum_{y_i^{(n)}=0}^{K} \pi(y_i^{(n)}) \frac{y_i^{(n)}}{K}.
\end{equation}
Note from (\ref{g+}) and Fig.~\ref{BDP} that the Wall probabilities $q_i^{(n)}$ do not result from a fixed-point solution, because they are directly expressed as a function of the Newsfeed probabilities $p_i^{(n)}$.

\subsection{Derivation of the balance equations}

Here we further develop and simplify the equations (\ref{p}) and (\ref{q}). We first consider the Markov chain associated with the Newsfeed of user $n=j\neq i$ (see top right part of Fig.~\ref{BDP}). To solve this birth-and-death process we define the quantity
\begin{equation}
\label{si}
\tau_j := \frac{\lambda^{(i)}\mathbf{1}_{\left\{i\in\mathcal{L}^{(j)}\right\}} + \sum_{k\in\mathcal{L}^{(j)}}\mu^{(k)}p_i^{(k)}}{\sum_{k\in\mathcal{L}^{(j)},k\neq i}\lambda^{(k)}+ \sum_{k\in\mathcal{L}^{(j)}}\mu^{(k)}(1-p_i^{(k)})}.\nonumber
\end{equation}
Note that $\tau_j$ depends on all probabilities $p_i^{(k)}$ for $k \in \mathcal{L}^{(j)}$. The steady-state probability of the Markov chain associated with the Newsfeed of user $j$ can then be derived using $\tau_j$ (see \cite[Section 1.3, eq.1.9]{Kelly} and transition rates in Fig.~\ref{BDP}):
\begin{equation}
\label{SSxi}
\pi(x_i^{(j)}) = \pi(0) \left(\begin{tabular}{c} $M$ \\ $x_i^{(j)}$ \end{tabular}\right)\tau_j^{x_i^{(j)}},\nonumber
\end{equation}
where $\pi(0)$ is obtained by normalization (thanks to the Binomial formula):
\begin{equation}
\pi(0) = \frac{1}{(1+\tau_j)^{M}}.\nonumber
\end{equation}
Then, applying this result in (\ref{p}) we get:
\begin{eqnarray}
\label{CFpj}
p_i^{(j)} & = & \frac{1}{(1+\tau_j)^{M}} \sum_{m=1}^{M}\left(\begin{tabular}{c} $M$ \\ $m$ \end{tabular}\right)\tau_j^m\frac{m}{M}\nonumber\\
& \stackrel{m':=m-1}{=} & \frac{\tau_j}{(1+\tau_j)^M}\sum_{m'=0}^{M-1}\left(\begin{tabular}{c} $M-1$ \\ $m'$ \end{tabular}\right)\tau_j^{m'}\nonumber\\
& \stackrel{Binomial}{=} & \frac{\tau_j}{1+\tau_j}. \nonumber
\end{eqnarray}
By replacing the expression for $\tau_j$, we obtain the following very simple expression for posts of origin $i$ in the Newsfeed of user $j$. The fixed-point is now visible:
\begin{equation}
\label{Cpj}
\left\lbrack \sum_{k\in \mathcal{L}^{(j)}}\left(\lambda^{(k)}+\mu^{(k)}\right) \right\rbrack p_i^{(j)} = \lambda^{(i)}\mathbf{1}_{\left\{i\in\mathcal{L}^{(j)}\right\}} + \sum_{k\in \mathcal{L}^{(j)}}\mu^{(k)}p_i^{(k)}.
\end{equation}
Following a similar reasoning, we get for the Newsfeed of any user $n= i$, the following equation: 
\begin{equation}
\label{Cpi}
\left\lbrack \sum_{k \in \mathcal{L}^{(i)}}\left(\lambda^{(k)}+\mu^{(k)}\right) \right\rbrack p_i^{(i)} = \sum_{k \in \mathcal{L}^{(i)}}\mu^{(k)}p_i^{(k)}.
\end{equation}

As a result, the set of $N$ equations (\ref{Cpj}) and (\ref{Cpi}) constitute the new equations of the fixed point, whose solution gives the required Newsfeed probabilities $p_i^{(n)}$ for $n=1,\ldots,N$.

In the same fashion, the steady-state probabilities for the Wall can be directly derived from the steady-state probabilities for the Newsfeed through the following equations:
\begin{eqnarray}
\label{Cqj}
\left(\lambda^{(j)}+\mu^{(j)}\right) q_i^{(j)} & = & \mu^{(j)} p_i^{(j)},\\
\label{Cqi}
\left(\lambda^{(i)}+\mu^{(i)}\right) q_i^{(i)} & = & \lambda^{(i)} + \mu^{(i)} p_i^{(i)}.
\end{eqnarray}

\subsection{Explanation of the balance equations}

Interestingly, equations (\ref{Cpj})-(\ref{Cpi}) and  (\ref{Cqj})-(\ref{Cqi}) allow for a simple intuitive interpretation: they balance the incoming and outgoing flow of posts of origin $i$ on each Newsfeed and Wall list. More precisely, equation (\ref{Cpj}) equalizes the incoming rate and the outgoing rate of posts of origin $i$ in the Newsfeed of user $j$ (for $j \neq i$). Here, $\sum_{k \in \mathcal{L}^{(j)}}\left(\lambda^{(k)}+\mu^{(k)}\right)$ is the average number of posts per unit of time that enter the Newsfeed of user $j$. From the random eviction policy, each of these arriving posts replaces a post of origin $i$ with probability $p_i^{(j)}$. 
Indeed, by assuming that post and re-post processes are Poisson, the PASTA property holds which tells us that arriving posts see the Newsfeed in steady-state.
As a result, the left-hand side of equation (\ref{Cpj}) is just the outgoing rate of posts of origin $i$ in the Newsfeed of user $j$. Now looking at the right-hand side of this equation, $\mu^{(k)}$ is the average number of posts per unit of time that arrive on the Newsfeed of user $j$ because a leader $k$ of $j$ reposts something on his/her Wall. Each of these posts is of origin $i$ with probability $p_i^{(k)}$, due to the random selection policy in Newsfeeds. In addition, if $i$ is a leader of $j$, the $\lambda^{(i)}$ self-posts of $i$ per unit of time also appear on the Newsfeed of $j$. As a result, the right-hand side of equation (\ref{Cpj}) is the incoming rate of posts  of origin $i$ in the Newsfeed of user $j$.

Similarly, equation (\ref{Cpi}) equalizes the incoming rate and the outgoing rate of posts of origin $i$ in the Newsfeed of user $i$. The only difference is that a new post that has just been created by $i$ does not appear on his/her own Newsfeed.

Equation (\ref{Cqj}) equalizes the incoming rate and the outgoing rate of posts of origin $i$ in the Wall of user $j$. Indeed $\lambda^{(j)}+\mu^{(j)}$ is the average number of posts per unit of time that enter the Wall of user $j$. Each of these posts replaces a post of origin $i$ with probability $q_i^{(j)}$, due to the random eviction policy in Walls and the PASTA property. As a result, the left-hand side of equation (\ref{Cqj}) is the outgoing rate of posts  of origin $i$ from the Wall of user $j$. Obviously, $\mu^{(j)} p_i^{(j)}$ is the average number of posts of origin $i$ per unit of time that arrive on the Wall of user $j$, due to the random selection policy in Newsfeeds.

Similarly, equation (\ref{Cqi}) equalizes the incoming and outgoing rate of posts of origin $i$ on the Wall of user $i$. We just have to add at the incoming rate the $\lambda^{(i)}$ self-posts per unit of time from $i$.

The balance equations involve the expected percentage of posts on the Newsfeeds and Walls of users and  have been derived based on an approximation (mean-field) on the rates of state transition. We prove in the following Theorem that, under some tighter assumptions, these equations are actually exact. However, note here that the detailed distribution of the number of posts on Newsfeeds and Walls $\pi(x)$, $\pi(y)$, from the birth-and-death processes in Fig.~\ref{BDP} are an approximation.
\begin{theo}[\textbf{Exactness}]
\label{exact}
For Poisson posting and re-posting activity with $\lambda^{(n)},\mu^{(n)}>0,\ \forall n\in\mathcal{N}$, strongly connected Follower graph and the random selection / random eviction policy, the equations (\ref{Cpj})-(\ref{Cpi}) and  (\ref{Cqj})-(\ref{Cqi}) describe exactly the original detailed model in steady-state.
\end{theo}
The proof can be found in Appendix B, and is based on \textit{the conservation law of posts} in the Newsfeed and in the Wall of each user.

The balance equations further give an important structural property of the steady-state solution as a side product.

\begin{cor}[\textbf{Insensitivity in list size}]
\label{Tins}
In view of (\ref{Cpj})-(\ref{Cpi}) and  (\ref{Cqj})-(\ref{Cqi}), the steady-state probabilities to find posts from user $i$ on the Newsfeed of any user $n$ ($p_i^{(n)}$, $n=1,\ldots,N$) as well as on the Wall of any user $n$ ($q_i^{(n)}$, $n=1,\ldots,N$), depend neither on the size $M$ of the Newsfeed, nor on the size $K$ of the Wall.
\end{cor}

\subsection{Alternative selection and eviction policies}
\label{pol}

Looking at the model through the balance equations enables us to relax random selection and random eviction assumptions, and introduce alternative policies. We will show now that a number of different policies satisfy the same balance equations, thus granting generality to the result.

\subsubsection{Random selection / FIFO eviction}

Let us modify the eviction policy in both Newsfeeds and Walls, and replace ``Random'' by a more realistic FIFO policy: now, a new post enters at the top of the list and evicts the oldest post out of the list.
To do so, we define $\phi^{(j)}_{i} $ as the new outgoing rate of posts of origin $i$ in the Newsfeed of user $j$. This will replace the left-hand side of eq. (\ref{Cpj})). From Little's law,
\begin{eqnarray}
\label{Xij23}
\phi^{(j)}_{i} & = &  \overline{X}^{(j)}_{i} / \overline{T}^{(j)}_{i},
\end{eqnarray}
where $\overline{X}^{(j)}_{i}$ is the average number of posts of origin $i$ in the Newsfeed of user $j$, and $\overline{T}^{(j)}_{i}$ is the average time a post of origin $i$ stays in the Newsfeed of user $j$.

The total arrival rate of posts in the Newsfeed of user $j$ is $\sum_{k \in \mathcal{L}^{(j)}}\left(\lambda^{(k)}+\mu^{(k)}\right)$. The mean time between two successive arrivals in the Newsfeed of user $j$ is thus the inverse of this quantity. As a result any post arriving in the Newsfeed of user $j$ will stay on average $M$ times this mean value:
\begin{equation}
\label{Tij}
\overline{T}^{(j)}_{i} = \frac{M}{\sum_{k \in \mathcal{L}^{(j)}}\left(\lambda^{(k)}+\mu^{(k)}\right)}.
\end{equation}
Since $\overline{X}^{(j)}_{i} = M p_i^{(j)}$ per definition, we conclude that:
\begin{eqnarray}
\label{Xij4}
\phi^{(j)}_{i} = p_i^{(j)} \sum_{k \in \mathcal{L}^{(j)}}\left(\lambda^{(k)}+\mu^{(k)}\right) & \mathrm{[FIFO\ evict]},
\end{eqnarray}
which has the same expression as the left-hand side of equation (\ref{Cpj}). In other words, when we replace the ``Random'' eviction policy by the FIFO eviction policy in the Newsfeed, equation (\ref{Cpj}) does not change. We can easily show by a similar reasoning that equations (\ref{Cpi}), (\ref{Cqj}) and (\ref{Cqi}) remain also unchanged under the FIFO eviction policy.

We now show that the set of balance equations remain the same also if we choose a TTL (Time-To-Live) eviction principle \cite{Har19}, \cite{GastHoudt17}. Here, each post stays at the Newsfeed for a fixed amount of time $T$ before leaving. In this case, the size of the list is not constant $M$, but rather varies over time. By Little's law, the mean Newsfeed size is equal to
\begin{eqnarray}
\label{sizeTTL}
\overline{M}^{(j)} & = & T\sum_{k\in\mathcal{L}^{(j)}}(\lambda^{(k)}+\mu^{(k)}),
\end{eqnarray}
where again $\sum_{k\in\mathcal{L}^{(j)}}(\lambda^{(k)}+\mu^{(k)})$ is the total arrival rate of posts in Newsfeed $j$. Then in (\ref{Xij23}) we substitute $\overline{X}_i^{(j)} = \overline{M}^{(j)}p_i^{(j)}$ and $\overline{T}_i^{(j)} = T$, to get 
\begin{eqnarray}
\phi_i^{(j)} = p_i^{(j)} \sum_{k \in \mathcal{L}^{(j)}}\left(\lambda^{(k)}+\mu^{(k)}\right) & \mathrm{[TTL\ evict]}.
\end{eqnarray}

\subsubsection{Newest selection / Random eviction}

We come back to our original model with a ``Random'' eviction policy, and where Newsfeeds are of limited size $M$ and Walls are of limited size $K$. We have proved in Theorem \ref{Tins} that the steady-state probabilities $p_i^{(j)}$ and $q_i^{(j)}$, being solutions of the system (\ref{Cpj})-(\ref{Cqi})) depend neither on $M$ nor on $K$. In order to change the selection policy from ``Random'' to ``Newest'', we just have to take $M = 1$. Indeed, when the size of Newsfeeds is unitary, the ``Random'' selection will necessarily choose the newest (i.e. latest, freshest) entry of the Newsfeed to repost on the Wall of a user. As a result, the system (\ref{Cpj})-(\ref{Cqi}) remains true also under the ``Newest'' selection policy.

Further extensions that incorporate user preferences towards posts of specific origins, or that consider user engagement metrics are very interesting topics for future research. Since the balance equations are derived based on the conservation law of posts on Newsfeeds (see Theorem~\ref{exact} and its proof in the Appendix) the latter can be the basis to analyse various alternative post selection and eviction policies.

%=======================================================================

\section{Closed Form solution}
\label{closed}

\subsection{Linear system}

We can re-write {(\ref{Cpj})-(\ref{Cpi}) and (\ref{Cqj})-(\ref{Cqi}) for posts with label $i$} in a compact form and summarize our findings as follows.

\begin{theo}[\textbf{Linear System}]
\label{th1}
The unknown column vectors $\mathbf{p}_i:=(p_i^{(1)},\ldots,p_i^{(N)})^T$ and $\mathbf{q}_i:=(q_i^{(1)},\ldots,q_i^{(N)})^T$ are the solution of the following linear system
\begin{eqnarray}
\label{LSa1}
\mathbf{p}_i & = & \mathbf{A}\cdot \mathbf{p}_i + \mathbf{b}_i\\
\label{LSb1}
\mathbf{q}_i & = & \mathbf{C}\cdot \mathbf{p}_i + \mathbf{d}_i.
\end{eqnarray}
\end{theo}

In the above, $\mathbf{A}$ and $\mathbf{C}$ are $N\times N$ matrices independent of $i$, whereas $\mathbf{b}_i$ and $\mathbf{d}_i$ are N-column vectors that depend on $i$. Hence, a standard linear system should be resolved for each $i$. The entries of the above matrices and vectors are summarised in Table \ref{T2}. It is interesting to note that $a_{j,j}=0$ for all $j$, $b_{i,i}=0$, $\mathbf{C}$ is diagonal, and also there is a unique positive $d_{j,i}$ entry for $i=j$. 

\begin{table}[ht!]
\caption{Entries for the matrices/vectors of the linear system.}
\centering
\begin{tabular}{|c|c|}
\hline
& \\
$\mathbf{A}$ & $a_{j,k}:=\frac{\mu^{(k)}}{\sum_{\ell\in\mathcal{L}^{(j)}}(\lambda^{(\ell)}+\mu^{(\ell)})}\mathbf{1}_{\left\{k\in\mathcal{L}^{(j)}\right\}}$\\
&  \\
\hline
&  \\
$\mathbf{b}_i$ & $b_{j,i} : = \frac{\lambda^{(i)}}{\sum_{\ell\in\mathcal{L}^{(j)}}(\lambda^{(\ell)}+\mu^{(\ell)})}\mathbf{1}_{\left\{i\in\mathcal{L}^{(j)}\right\}}$\\
&  \\
\hline
& \\
$\mathbf{C}$ & $c_{j,k}:=\frac{\mu^{(j)}}{\lambda^{(j)}+\mu^{(j)}}\mathbf{1}_{\left\{j=k\right\}}$\\
&  \\
\hline
 &  \\
$\mathbf{d}_i$ & $d_{j,i}:=\frac{\lambda^{(i)}}{\lambda^{(i)}+\mu^{(i)}}\mathbf{1}_{\left\{j=i\right\}}$\\
& \\
\hline
\end{tabular}
\label{T2}
\end{table}

The matrix $\mathbf{A}$ is non-negative. In addition, it is row sub-stochastic, meaning that the sum of all its rows is less than or equal to $1$, with at least one row sum strictly less than $1$ (if we reasonably assume that at least one user injects self-posts). Another interesting property is that $\mathbf{A}$ is a weighted version of the Follower-matrix $\mathbf{F}=\mathbf{L}^T$, so that if $\mathbf{1}_{\left\{j\in\mathcal{F}^{(k)}\right\}} = \mathbf{1}_{\left\{k\in\mathcal{L}^{(j)}\right\}} = 0 \Rightarrow a_{j,k}=0$. There are cases however where $j$ follows $\ell$, but $a_{j,\ell}=0$ in the matrix $\mathbf{A}$, because $\mu^{(\ell)}=0$. Hence, users that never re-post from their Newsfeed alter the possibilities of post propagation in the graph. This is why we call $\mathbf{A}$, the \textit{propagation matrix}.

%%%%%%%

\subsection{Closed-form solution}

\begin{theo}[\textbf{Solution}]
\label{Main}
If the spectral radius $\rho(\mathbf{A})<1$, then the solution of the linear system (\ref{LSa1})-(\ref{LSb1}) is unique, %and given by
\begin{eqnarray}
\label{Th1a}
\mathbf{p}_i & = & \left(\mathbf{I}_N-\mathbf{A}\right)^{-1}\mathbf{b}_i\\
\label{Th1b}
\mathbf{q}_i & = & \mathbf{C} \left(\mathbf{I}_N-\mathbf{A}\right)^{-1}\mathbf{b}_i + \mathbf{d}_i.
\end{eqnarray}

\end{theo}
\begin{proof}
This results directly from \cite[Chapter 6, Lemma 2.1]{BerPleNN}, that we also include in Appendix C.
\end{proof}

An interesting observation is that the inverse $(\mathbf{I}_N-\mathbf{A})^{-1}$ involved in the derivation of $\mathbf{p}_i$ (relation~(\ref{Th1a})) is independent of $i$. Thus, in the solution process the inverse should be calculated only once, and then applied to the expressions in (\ref{Th1a})-(\ref{Th1b}) for labels $i=1,\ldots,N$.

We would like to know under which conditions a solution to the linear system exists, in other words when does $\rho(\mathbf{A})<1$ holds true, based on the specific structure of the non-negative matrix $\mathbf{A}$ given in Table~\ref{T2}. We show the following property.

\begin{lem}
\label{Lemma2}
It holds $\rho(\mathbf{A})\leq 1$. Strict inequality is guaranteed in the following non-exclusive non-exhaustive cases (cs):
\begin{enumerate}[(cs1)]
\item $\lambda^{(n)}>0$, $\forall n\in\mathcal{N}$.
\item For every cycle in the Leader-graph, at least one participating user has a leader $k$ with positive self-post rate.
\end{enumerate}
\end{lem}

\begin{proof}
Let us denote the row~sums of $\mathbf{A}$ by $r(j)$, $j=1\ldots N$. Then $r(j)\leq 1$ by definition from Table \ref{T2}. It is known that (\cite[Theorem 8.1.22]{HornJohn}) the following bounds are valid for the spectral radius of a non-negative matrix: $\min_{j=1}^N r(j) \leq \rho(\mathbf{A}) \leq \max_{j=1}^N r(j)$. The right-hand side in our case is $1$ and the first part is proven. 

(cs1) When $\lambda^{(n)}>0$, $\forall n$, then $\forall j$ and $k\in\mathcal{L}^{(j)}$, $a_{j,k}< \mu^{(k)} / \sum_{\ell\in\mathcal{L}^{(j)}}\mu^{(\ell)}$, so that $r(j)<1$, $\forall j$. Then the matrix is \textit{strictly} sub-stochastic, and $\rho(\mathbf{A})\leq \max_{j=1}^N r(j)<1$.

(cs2) In this case, suppose the length of a particular cycle is $\gamma>1$ and the participating nodes are $n_1,\ldots,n_{\gamma}$. Then at least one row sum $r(j)<1$, $j\in\left\{n_1,\ldots,n_{\gamma}\right\}$. By direct application of the Al'pin, Elsner, van den Dreissche bound \cite[Theorem A]{EvD08}, we conclude that $\rho(\mathbf{A})<1$. An additional condition for this bound is that $r(j)>0$, $\forall j$, which is satisfied when $\mathcal{L}^{(j)}\neq \emptyset$, $\forall j\in\mathcal{N}$ and not all leaders of some user have $\mu^{(k)}=0$.
\end{proof}

\begin{rem}
A special instance of (cs2) is when $\mathbf{A}$ is irreducible and $\lambda^{(j)}>0$ for at least one $j\in\mathcal{N}$.
\end{rem}

\subsection{Fixed-point algorithm}

For large $N$ it can be practically very difficult to calculate the inverse $\left(\mathbf{I}_N-\mathbf{A}\right)^{-1}$. A different way to proceed in order to solve the system (\ref{LSa1}) is to use an iterative approach.

\begin{theo}
\label{th3}
For the two cases of Lemma \ref{Lemma2} and any initialization vector $\mathbf{p}_i(0)$, the discrete-time linear system (\ref{LSa}) converges towards the fixed-point solution (\ref{Th1a}) when $t\rightarrow \infty$.
\begin{equation}
\label{LSa}
\mathbf{p}_i(t) = \mathbf{A}\cdot \mathbf{p}_i(t-1) + \mathbf{b}_i.
\end{equation}
The rate of convergence is the spectral radius of $\mathbf{A}$, $\rho(\mathbf{A})$,
\begin{eqnarray}
\label{UBrho}
 \min_{j=1\ldots N} r(j) \leq \rho(\mathbf{A}) \leq \max_{j=1\ldots N} r(j),
\end{eqnarray}
where $r(j)=\sum\limits_{\ell\in\mathcal{L}^{(j)}}\mu^{(\ell)}/\sum\limits_{\ell\in\mathcal{L}^{(j)}}(\lambda^{(\ell)}+\mu^{(\ell)})$. In the homogeneous case where $\mu^{(\ell)}=\mu, \lambda^{(\ell)}=\lambda$ $\forall \ell$, it holds $\rho(\mathbf{A}) = \frac{\mu}{\lambda+\mu}$.
\end{theo}

\begin{proof} We first write $\mathbf{p}_i(t)$ as a function of $\mathbf{p}_i(0)$ and $t$,
\begin{equation}
\label{Pit1}
\mathbf{p}_i(t) = \mathbf{A}^{t}\mathbf{p}_i(0) + \left(\sum_{n=0}^{t-1}\mathbf{A}^n\right)\mathbf{b}_i.\nonumber
\end{equation}
We need to find the limiting value $\mathbf{p}_i:=\lim_{t\rightarrow\infty}\mathbf{p}_i(t)$. For the two cases in Lemma \ref{Lemma2} we have $\rho(\mathbf{A})<1$, so that from \cite[pp.137--138, or Theorem 5.6.12]{HornJohn} it holds $\mathbf{A}^{\infty}:=\lim_{t\rightarrow\infty}\mathbf{A}^{t} = \mathbf{0}$. Additionally, from  \cite[Chapter 6, Lemma 2.1]{BerPleNN} (see Lemma \ref{Lemma1}) the limit of the matrix series for $t\rightarrow\infty$ converges to $\left(\mathbf{I}_N-\mathbf{A}\right)^{-1}$. Hence, the iteration converges to the solution (\ref{Th1a}), and is independent of the initialisation $\mathbf{p}_i(0)$. 

The error is defined as $\epsilon(t) = \mathbf{p}_i(t)-\mathbf{p}_i$, where $\mathbf{p_i}$ solves (\ref{LSa1}). Then from (\ref{LSa}) we get $\epsilon(t) = \mathbf{A}^t\epsilon(0)$, which tends to zero with rate that is dominated by the largest eigenvalue. From (\cite[Theorem 8.1.22]{HornJohn}) $\min_{j=1}^N r(j) \leq \rho(\mathbf{A}) \leq \max_{j=1}^N r(j)$, where $r(j)$ is the sum of row $j$.

\end{proof}

Note that once the Newsfeed-vector $\mathbf{p}_i:=\lim_{t\rightarrow\infty}\mathbf{p}_i(t)$ has been obtained, the Wall-vector $\mathbf{q}_i$ can be calculated from relation~(\ref{LSb1}). The influence score for user $i$, i.e. the value $\Psi_i$ is then directly derived from (\ref{aim1}). We need to solve for all $i$'s to derive all scores $\{\Psi_i\}_{i=1}^N$, however notice that the matrices $\mathbf{A}$ and $\mathbf{C}$ are the same for all users, and only $\mathbf{b}_i$ and $\mathbf{d}_i$ differ.

%%%%%%%%%%%%%%PageRank %%%%%%%%%%%%%%%%%%%%%%%%%%%%
\section{Example and Relation with PageRank}
\label{PageRank}
\begin{figure}[t!]
\centering
\hspace{-.cm}
	%\subfloat[Toy Graph]
	{\includegraphics[scale=0.2]{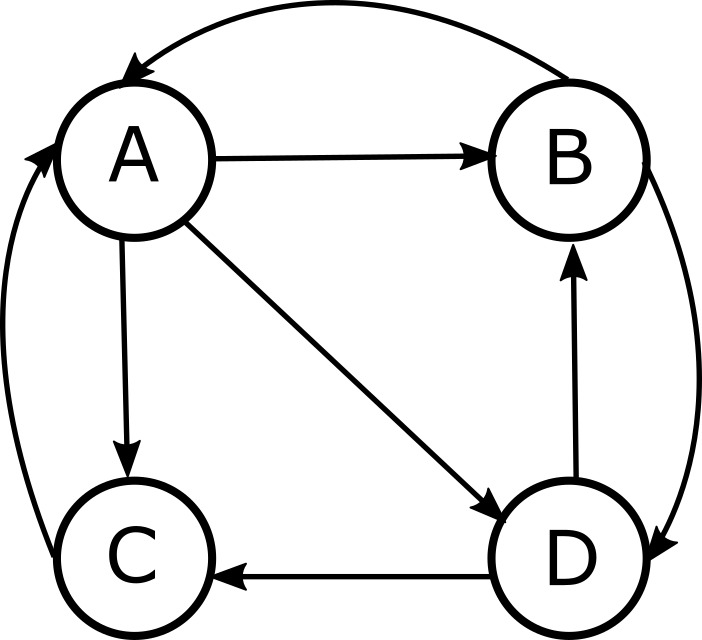}\label{toyG}} \hspace{0.6cm}
	%\subfloat[Follower graph]
	{\includegraphics[scale=0.3]{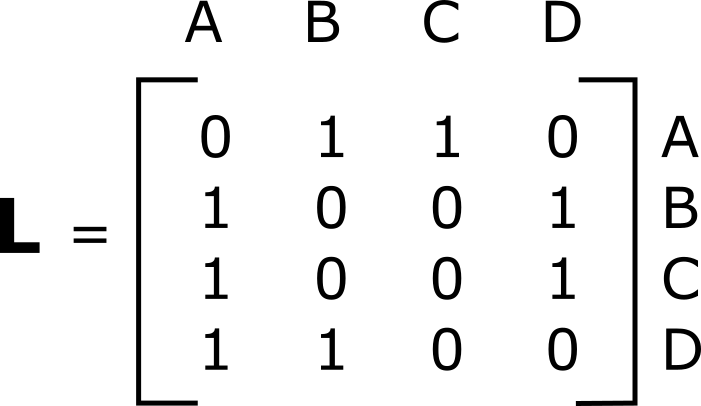}\label{toyM}} \hspace{0.5cm}
\caption{Example graph with adjacency matrix: the edge $j\rightarrow i$ here means that $j$ refers to $i$ (PageRank) or that $j$ follows $i$, i.e. $\mathbf{1}_{\left\{j\in\mathcal{F}^{(i)}\right\}}=\mathbf{1}_{\left\{i\in\mathcal{L}^{(j)}\right\}}=1$  (Newsfeed), and we set $\ell_{i,j}=1$ in the leader matrix $\mathbf{L}$.}
\label{fig:toy}
\end{figure}

In this section we show and prove that PageRank  \cite{pagerank98} is a special case of the $\tilde{\Psi}$-score variant in (\ref{prQ}), for uniform activity of all users. As an illustrative example, we use the toy-graph in Fig.~\ref{fig:toy}, which is strongly connected. For notation, column-vector $\mathbf{e}$ has $1$'s in all entries, and column-vector $\mathbf{e}_i$ has all entries $0$, except position $i$ with value $1$.\\
%%%%%%%%%%%%%% Plots de validation %%%%%%%%%%%%%%%%%%%%%%%%
%%%%%%%%%%%%%%%%%%%%%%%%%%%%%%%%%%%%%%%%%%%%%%%%
\textit{PageRank:} The famous PageRank score $\pi$ (column-vector) is calculated \cite{ScaleRank} based only on the graph topology,
\begin{eqnarray}
\label{PR1}
\mathbf{\pi} & = &  \beta \mathbf{W}\mathbf{\pi} +(1-\beta)\frac{\mathbf{e}}{N}.
\end{eqnarray}
Here, $\mathbf{W}=\mathbf{L}\mathbf{D}_{out}^{-1}$, where $\mathbf{D}_{out}$ is a diagonal matrix with non-zero entries the out-degrees of the nodes $[deg_{out}^{(1)},\ldots,deg_{out}^{(N)}]$. Its inverse multiplies on the right the leader-adjacency matrix $\mathbf{L}$, so that the matrix $\mathbf{L}\mathbf{D}_{out}^{-1}$ is column-stochastic, i.e. the sum of all columns is $1$. For the toy example in Fig.~\ref{fig:toy},
\begin{eqnarray}
\mathbf{W}& = & \left[
\begin{tabular}{c c c c}
$0$ & $1/2$ & $1$ & $0$\\
$1/3$ & $0$ & $0$ & $1/2$\\
$1/3$ & $0$ & $0$ & $1/2$\\
$1/3$ & $1/2$ & $0$ & $0$
\end{tabular}
\right].\nonumber
\end{eqnarray}
With probability $(1-\beta)$ the random surfer teleports from his/her current state, to any other state chosen uniformly at random. This trick is useful to avoid dead-ends. We can solve iteratively (\ref{PR1}) to find the PageRanks $[\pi_A,\pi_B,\pi_C,\pi_D] = [0.331,0.223,0.223,0.223]$, whose sum equals $1$. We observe that node $A$ has highest PageRank, whereas the other three nodes have equal PageRank.

\textit{$\Psi$-score:} For the sake of comparison, we let all $\lambda^{(i)}=\lambda$ and $\mu^{(i)}=\mu$, for $i=1,\ldots,N$. Furthermore, let us choose values of $\mu$ and $\lambda$ such that $\beta=\mu/(\lambda+\mu)$. In this homogeneous case, only the graph topology is important and the Newsfeed system of equations in (\ref{LSa1}) is written for row-vector $\mathbf{p}_i^T$ as
\begin{eqnarray}
\label{Pside}
\mathbf{p}_i^T & = & \beta\mathbf{p}_i^T\mathbf{W} + (1-\beta)\mathbf{W}_{(i,:)}, \ \ \ \ \forall i\in\mathcal{N}.
\end{eqnarray}
Here, the unknown row vector multiplies $\mathbf{W}$ from the left. Note that $\mathbf{A}= \beta\mathbf{W}^T$ in (\ref{LSa1}). Furthermore, the teleportation is towards the row-vector $\mathbf{W}_{(i,:)}$, where $\mathbf{W}_{(i,:)}$ is the $i$-th row of the matrix $\mathbf{W}$. There are $N$ such systems of equations (\ref{Pside}), one for each node $i\in\left\{A,B,C,D\right\}$. For each user $i$, we can calculate his/her influence column-vector $\mathbf{p}_i$ on the Newsfeed of every other user, by power iteration as in (\ref{LSa}), and gather our findings in matrix $\mathbf{P}$. 
Having calculated $\mathbf{p}_i$ for each node, we replace in (\ref{LSb1}), which in our homogeneous case becomes
\begin{eqnarray}
\label{Qside}
\mathbf{q}_i & = & \beta\mathbf{p}_i + (1-\beta)\mathbf{e}_i,
\end{eqnarray}
so we get the influence of $i$ on every Wall in the network. All column-vectors of Wall-influence are grouped in matrix $\mathbf{Q}$.
Using the $\tilde{\Psi}$-score definition in (\ref{prQ}), $\tilde{\Psi}_i=\sum_{n=1}^Nq_i^{(n)}/N$ for origin $i$, we average over all entries of column $i$ in the $\mathbf{Q}$ matrix including self-influence at the diagonal. Repeating for $i\in\left\{A,B,C,D\right\}$ we get $\left[\tilde{\Psi}_A,\tilde{\Psi}_B,\tilde{\Psi}_C,\tilde{\Psi}_D\right] = \left[0.331, 0.223, 0.223,0.223\right]$, equal to the PageRank $\pi$ score.

\begin{theo}
\label{pagerankQ}
In the homogeneous case of uniform user activity with $\lambda^{(n)}=\lambda$ and $\mu^{(n)}=\mu\ $, $\forall n\in\mathcal{N}$, the score $\tilde{\Psi}_i = \sum_{n=1}^Nq_i^{(n)}/N$ from (\ref{prQ}) is PageRank $\pi_i$ with damping factor $\beta=\frac{\mu}{\lambda+\mu}\in(0,1)$ (teleportation $1-\beta=\frac{\lambda}{\lambda+\mu}$).
\end{theo}
\begin{proof}
To show this, let us first write the homogeneous activity Newsfeed and Wall equations of all users in matrix form, using the influence matrices $\mathbf{P}$ and $\mathbf{Q}$. From (\ref{Pside}) and (\ref{Qside})
\begin{eqnarray}
\label{P}
\mathbf{P}^T & = & \beta\mathbf{P}^T\mathbf{W} + (1-\beta)\mathbf{W},\\
\label{Q}
\mathbf{Q}^T & = & \beta\mathbf{P}^T + (1-\beta)\mathbf{I},
\end{eqnarray}
where as discussed $\mathbf{W}:=\beta^{-1}\mathbf{A}^T$ from Table~\ref{T2}.
Using (\ref{Q}) we can derive the score column-vector $\mathbf{\tilde{\Psi}}$ defined in (\ref{prQ})
\begin{eqnarray}
\label{PsiEq}
\mathbf{\tilde{\Psi}} = \frac{1}{N}\mathbf{Q}^T\mathbf{e} & = & \beta\frac{1}{N}\mathbf{P}^T\mathbf{e} + (1-\beta)\frac{1}{N}\mathbf{e}.
\end{eqnarray}
We solve (\ref{P}) over $\mathbf{P}^T$ and do the following manipulation
\begin{eqnarray}
\label{manipP}
\mathbf{P}^T & = &  (1-\beta)\mathbf{W}(\mathbf{I} - \beta\mathbf{W})^{-1}\nonumber\\
& \stackrel{(*)}{=} & (1-\beta)\mathbf{W}\sum_{t=0}^{\infty}(\beta\mathbf{W})^{t} =  \frac{1-\beta}{\beta}\sum_{t=1}^{\infty}(\beta\mathbf{W})^{t}\nonumber\\ 
& = & \frac{1-\beta}{\beta}\left((\mathbf{I} - \beta\mathbf{W})^{-1} - \mathbf{I}\right),
\end{eqnarray}
where (*) uses Lemma~\ref{Lemma1} in Appendix, because $\beta\mathbf{W}$ (the matrix $\mathbf{A}^T$) is column sub-stochastic. Replacing the above in (\ref{PsiEq})
\begin{eqnarray}
\label{PsiEq2}
\mathbf{\tilde{\Psi}} & = & (1-\beta)\frac{1}{N}\left((\mathbf{I} - \beta\mathbf{W})^{-1} - \mathbf{I}\right)\mathbf{e} + (1-\beta)\frac{1}{N}\mathbf{e}.\nonumber\\
& = &  (1-\beta)(\mathbf{I} - \beta\mathbf{W})^{-1}\mathbf{e} \frac{1}{N}\Rightarrow\nonumber\\
\mathbf{\tilde{\Psi}} & = & \beta\mathbf{W}\mathbf{\tilde{\Psi}} + (1-\beta)\mathbf{e} \frac{1}{N}.
\end{eqnarray}
The last equation is the same as PageRank in (\ref{PR1}).
\end{proof}

Since activity is homogeneous here, then $\mu/(\lambda+\mu)$ is the probability that some user reposts and $\lambda/(\lambda+\mu)$ the probability that some user posts. Then the score of a user is fed from the reposts of its direct followers and its own self-posts.

The greatest powers of our method and the $\Psi$-score, however, are revealed when ranking users with asymmetric activity. \\
$\bullet$ Suppose a scenario with the same toy-graph and activities $\mu=2$ for all users re-posting, but $\lambda_A=\lambda_B=\lambda_D=\lambda=0.105$, $\lambda_C=3\lambda=0.315$ for posting. Then $\lambda,\mu$ are again such that $\mu/(\lambda+\mu)=\beta=0.95$, but user $C$ posts with $3$-times higher frequency than the others. After solving the system (\ref{LSa1})-(\ref{LSb1}) for these new parameter values, we can calculate the new $\Psi$-score using (\ref{prQ}), to get  $\left[\tilde{\Psi}'_A,\tilde{\Psi}'_B,\tilde{\Psi}'_C,\tilde{\Psi}'_D\right] = \left[0.234,0.156,0.451,0.159\right]$. User $C$ is now ranked first with users $A$, $B$ and $D$ following; our score takes the increased posting activity into account.\\
$\bullet$ In another scenario with the same toy-graph and activities $\mu_A=\mu_B=\mu_D=\mu=2$, $\mu_C=0$, and $\lambda=0.105$ for all users, user $C$ decides to stop re-posting anything. We can again solve the system  (\ref{LSa1})-(\ref{LSb1}) and get the new scores $\left[\tilde{\Psi}''_A,\tilde{\Psi}''_B,\tilde{\Psi}''_C,\tilde{\Psi}''_D\right] = \left[0.122, 0.231, 0.468,0.179\right]$. This behaviour of user $C$ results again in him/her ranking higher than the other users, by not sharing anything on his/her Wall.

From the above examples, we can safely conclude that the $\Psi$-score is much more expressive than PageRank: it incorporates the user posting and reposting activity in an appropriate way in the score. This is shown more emphatically in Section~\ref{trace} for real world traces, where the user $\Psi$-scores are quite different from other scores in practice (e.g. PageRank, \#Followers). For its calculation one derives the $\mathbf{p}_i$ and $\mathbf{q}_i$ vectors, which contain detailed information of the influence of $i$ on any Newsfeed and Wall in the network. To obtain such fine grained information, we pay in complexity; to derive the $\Psi_i$-score for user $i$ using Theorem~\ref{th3} every iteration in (\ref{LSa}) includes a weighted matrix-vector multiplication plus a vector addition to find $\mathbf{p}_i$. Such iteration is very similar to the power method that calculates the PageRanks. Since we need to solve the system for every user $i=1,\ldots,N$, to calculate all the $\Psi$-scores requires $N$-times the complexity that PageRank would need. Other algorithms with lower complexity or faster convergence that work for PageRank (see e.g.\cite{ScaleRank}) could also apply for the $\Psi$-score. This is a very interesting topic for future investigations.

%%%%%%%%%%%%%% Algorithm %%%%%%%%%%%%%%%%%%%%%%%%%%%%

\section{Implementation and Numerical Aspects}
\label{algo1}

For the numerical implementation we coded the following programs, that we make available in \cite{code}: (A) an algorithm to derive the $\Psi$-rank for each user from the balance equations of the model; two versions are coded, one for small OSP sizes and a sparse one for real-world sizes, (B) a discrete-event simulator, to simulate over time the behaviour of an OSP with arbitrary input traffic and user/platform policies, and (C) an emulator, which takes a real data-trace as input and outputs empirical $\Psi$-scores. 

\subsection{$\Psi$-ranking by the model}
\label{algo}

We remind the reader that the $\Psi$-score of user $i$'s influence in the social platform was introduced in (\ref{aim1}), as a function of the $q_i^{(j)}$'s ($j=1,\ldots,N$), i.e. the steady-state Wall probabilities. These values have been derived in closed form, through the balance equations. The two methods to calculate them, one using Theorem \ref{Main} with matrix inversion and a second using Theorem \ref{th3}, are coded in \cite{code} for small OSP sizes. The algorithm takes as input the vector of all posting and reposting rates $\lambda=\left(\lambda_1,\ldots,\lambda_N\right)$ and $\mu=\left(\mu_1,\ldots,\mu_N\right)$, as well as the graph $\mathcal{G}$ and outputs the $\Psi$-scores. 

Since the size $N$ of real-world social graphs is of the order of millions of users, an efficient algorithm that runs in reasonable time-scales is necessary to calculate these scores for all users in the platform. A tedious matrix inversion $\left(\mathbf{I}_N-\mathbf{A}\right)^{-1}$ is not recommended for such cases, as it is computationally very expensive - typically of the order of $\mathcal{O}(N^3)$. We have programmed here another algorithm to calculate the $\Psi$-score in large graphs, which implements a sparse version of the iterative solution introduced in Theorem \ref{th3}. It computes first the $\mathbf{p}_i$'s by involving at each iteration a matrix-vector multiplication $\mathbf{A}\cdot \mathbf{p}_i$ and then a vector addition $\mathbf{b}_i$, which take into account the sparsity of both the propagation matrix $\mathbf{A}$, and each vector $\mathbf{b}_i$. Sparsity comes from the fact that the number of leaders and followers of any user $i$ is very small compared to the total population $N$; as a second step - to calculate the $\mathbf{q}_i$'s we also use the fact that $\mathbf{C}$ is diagonal and sparse, and $\mathbf{d}_{i}$ has a single non-zero element. We can break the user set in subsets of users and parallelise the computational process on several machines, because solving for the influence of one user $i$ does not affect the process of solving for others.

\subsection{Discrete-event simulator}
\label{simu}

Additionally, we have developed our own discrete-event simulator (also available in \cite{code}) to validate the mathematical analysis through simulation, and to evaluate the robustness of the modelling assumptions presented in Section \ref{assumptions} against alternative traffic and policies. Unlike the code in the previous section \ref{algo} which solves the model's balance equations, the simulator precisely implements the behaviour of the generic OSP over time as described in Section~\ref{systemd}: i) The global state description consists of dynamic lists (of length $K$ for Walls and $M$ for Newsfeeds); ii) A variety of selection and eviction policies are implemented (``Random'', in a first phase, and ``Newest'', FIFO, ``Popular'' later to evaluate robustness); iii) Self- and re-posts can be generated according to Poisson or other processes. As such, the simulator \textit{does not} decouple the state space, \textit{does not} estimate average probabilities, and \textit{does not} rely on Markovian assumptions.
For each simulation we set $M=20$ and $K=10$ and ran long enough simulations to reach the steady-state with small confidence intervals. More specifically, in all experiments, we let the simulator run for a total of $\numprint{300000}$ events (self- and re-posts). 

\subsection{Emulator (Trace-based empirical influence)}
\label{emuli}
Finally, we have coded the emulator, which uses a real data-trace as input from Twitter, Weibo or other platform. The emulator differs from the simulator in the sense that it does not simulate the post propagation or specific policies, rather it directly outputs numerical values of influence, as read from the post sequence in the trace. To do so, we first pre-process the available data trace, so that each line of the input is just the quadruple $[\mathrm{PostID,\ TimeStamp,\ UserID,\ RePostID}]$, where the fourth entry is $-1$ in case of an original post, else the $\mathrm{PostID}$ of the original post which was reposted. The program calculates the influence of each user $i$ on another user $j$ directly from data ${q}^{emu}$[i][j], as the percentage of time that posts with origin $i$ are found on the Wall of user $j$. With these empirical values we derive the empirical score $\Psi^{\mathrm{emu}}_i$ for all users in the trace. The empirical influence is determined by the Wall occupancy periods divided by the total duration of the data-trace, and no further information about the social graph is necessary. We provide the emulator code in \cite{code}.

%%%%%%%%%%%%%% Plots de perf %%%%%%%%%%%%%%%%%%%%%%%%
\begin{figure*}[th!]
\centering
	\subfloat{\includegraphics[scale=0.3]{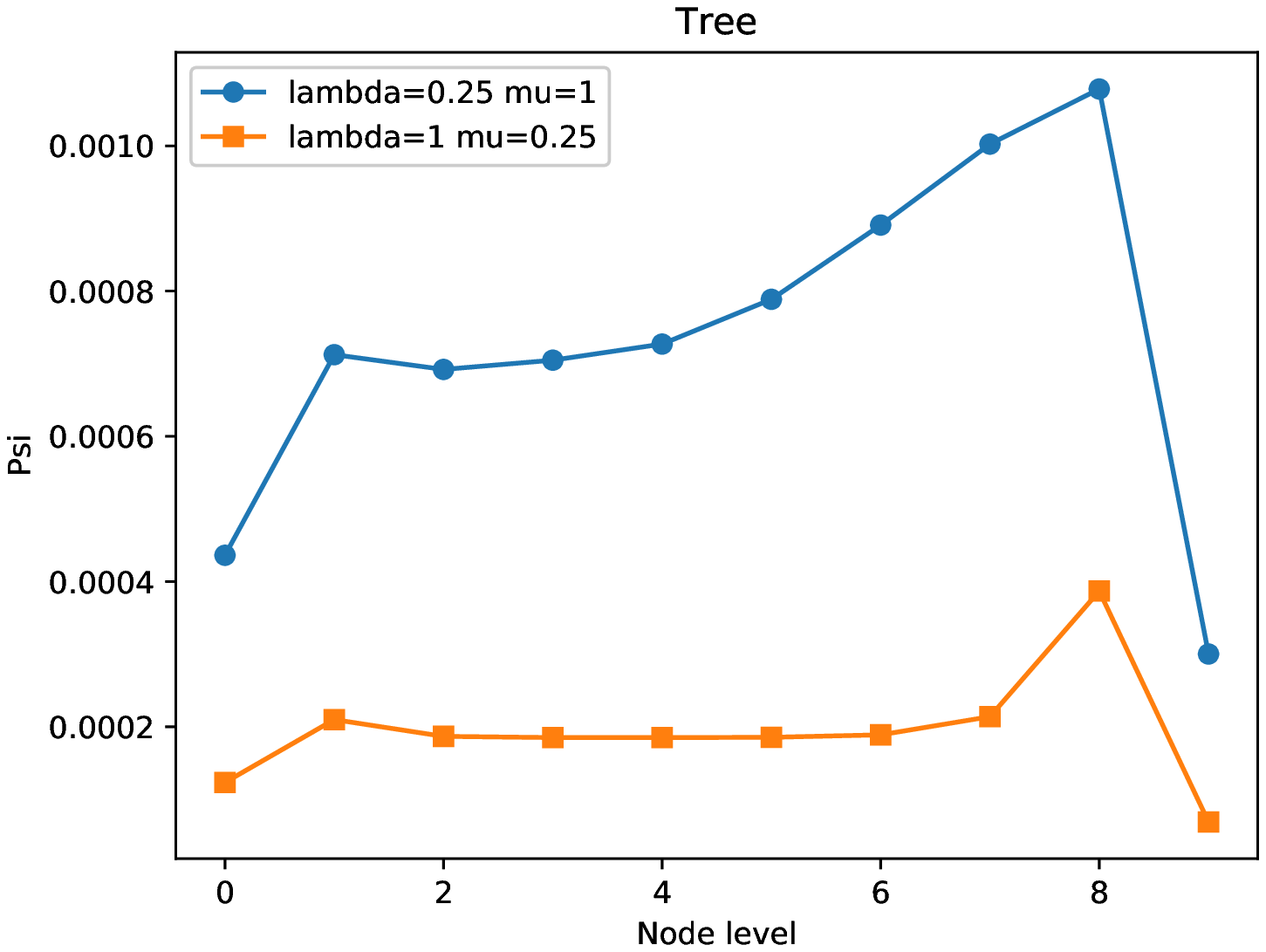}\label{tree_pos}} 
	\subfloat{\includegraphics[scale=0.3]{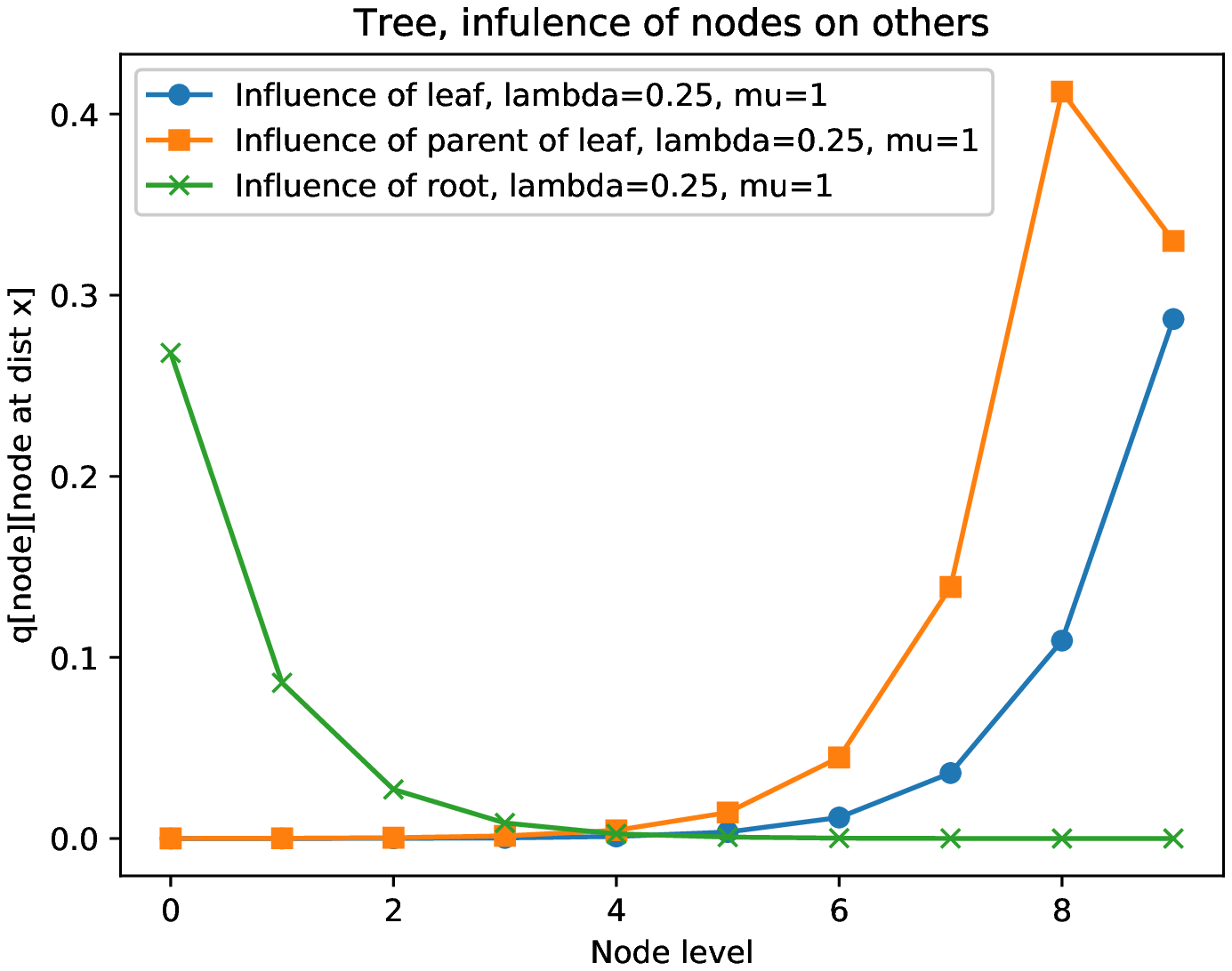}\label{tree_q}}
	\subfloat{\includegraphics[scale=0.3]{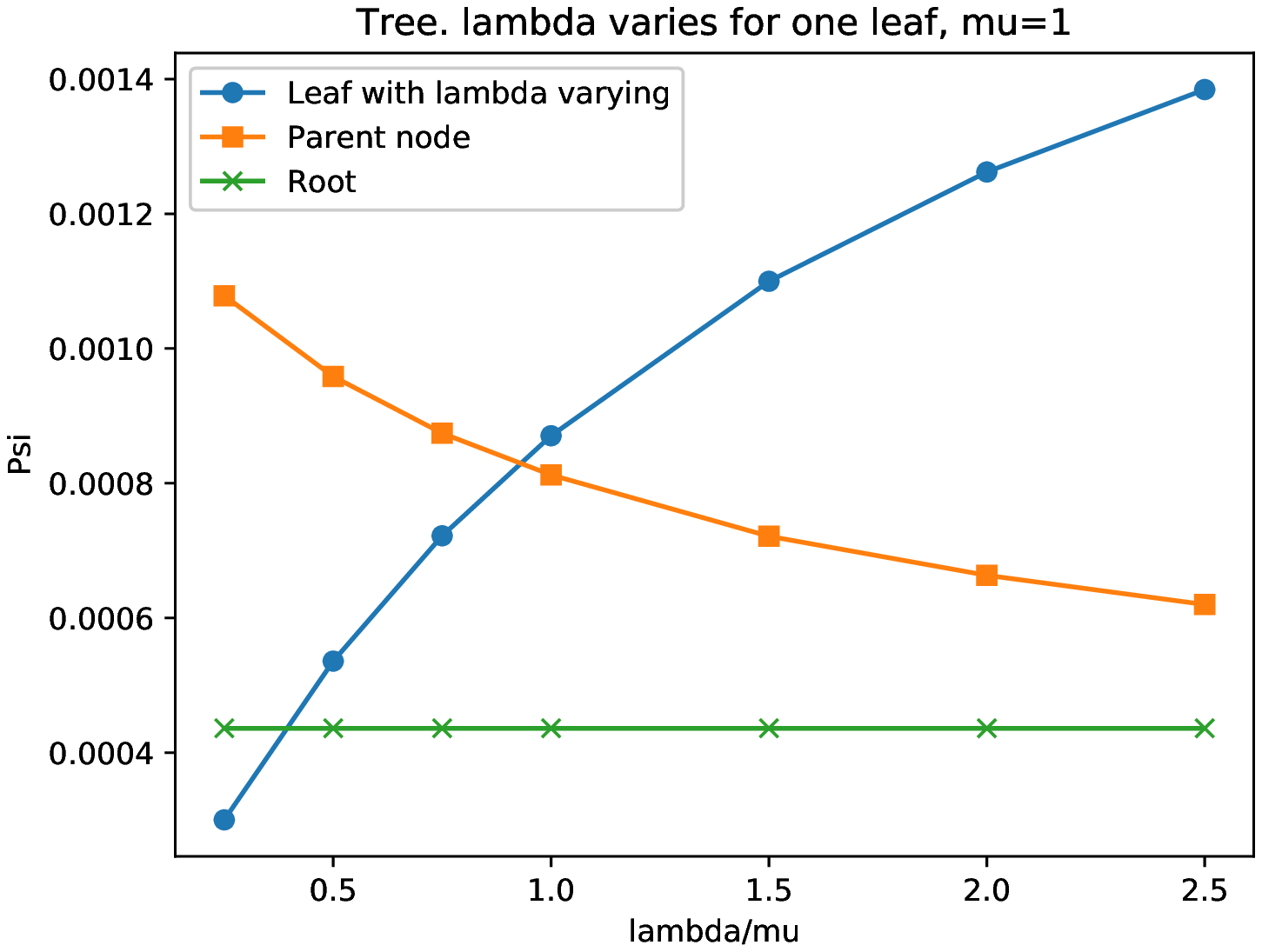}\label{tree_lam}}\\
	\subfloat{\includegraphics[scale=0.3]{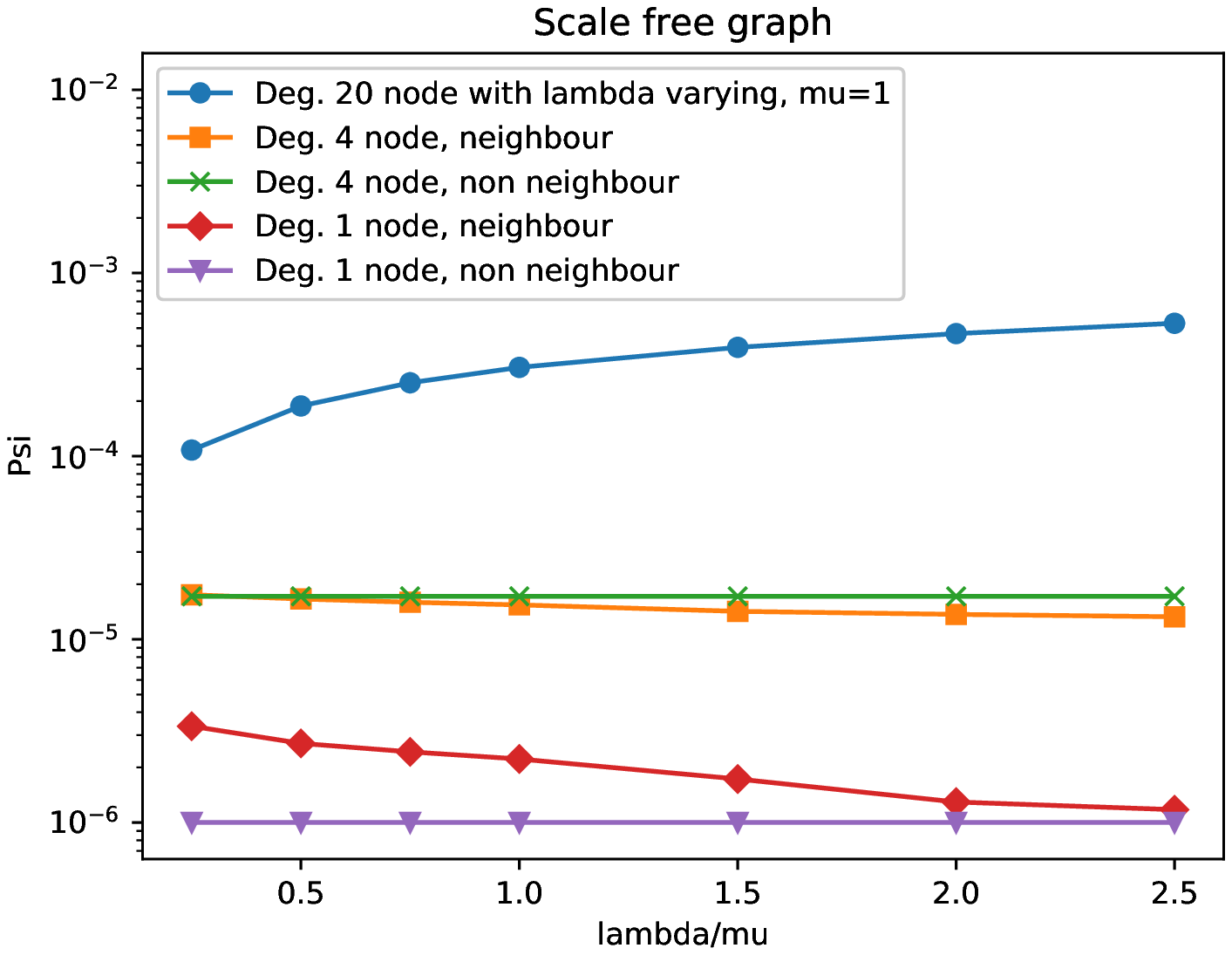}\label{sf_lam}} 
	\subfloat{\includegraphics[scale=0.3]{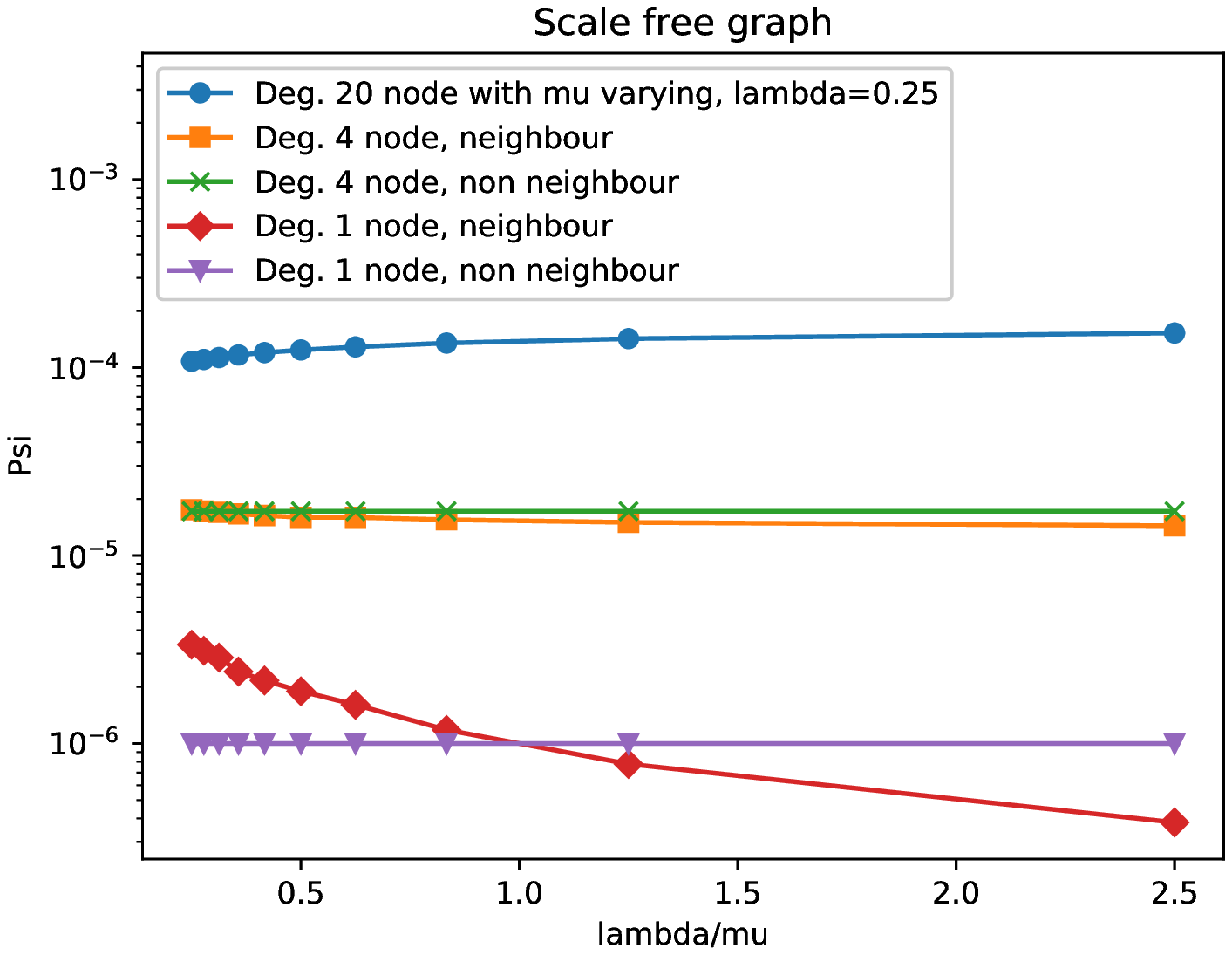}\label{sf_mu}}
	\subfloat{\includegraphics[scale=0.3]{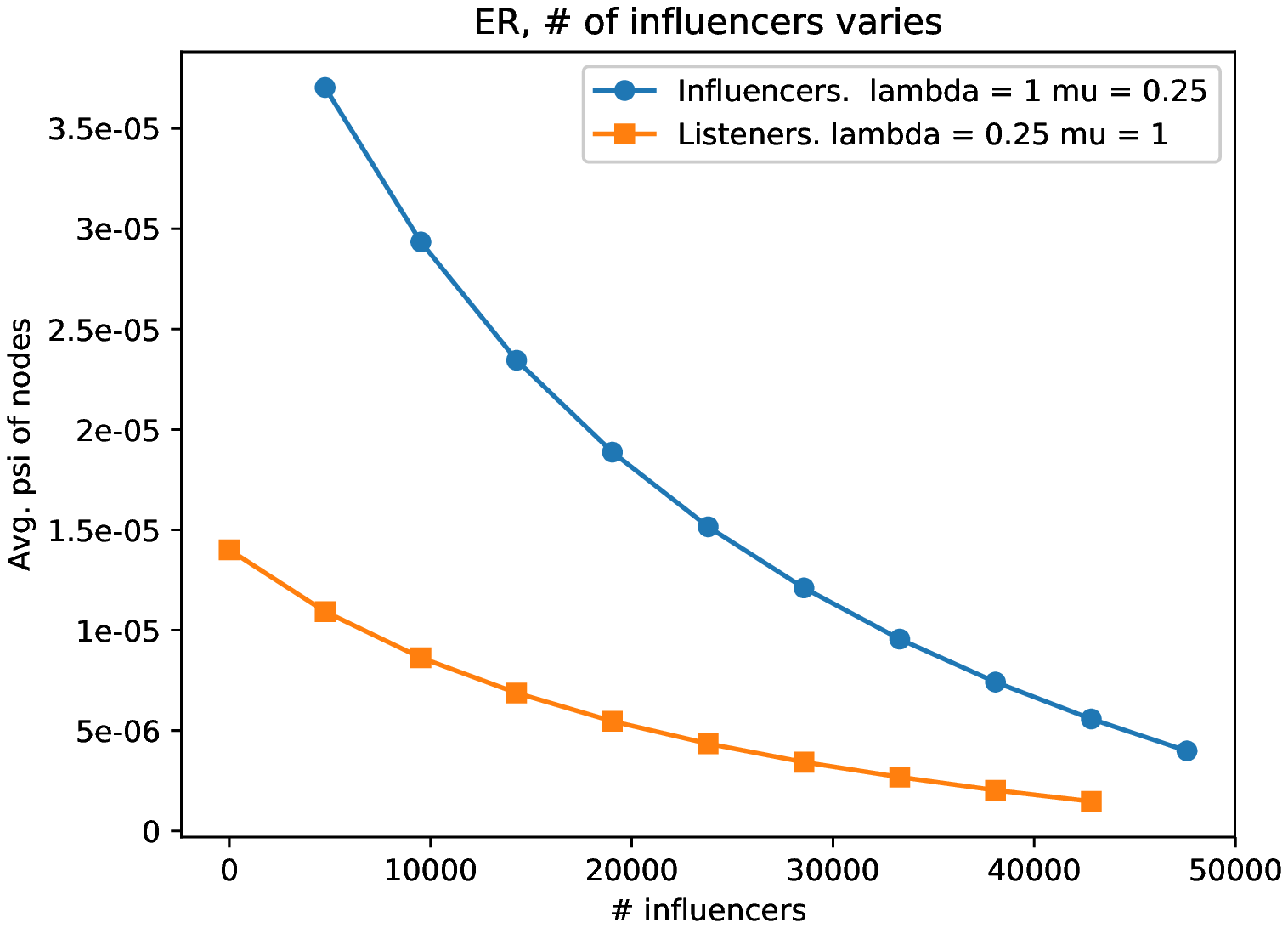}\label{er_tragedy}}
\caption{\textbf{Top:} Binary Tree. (a. left) $\Psi$-score per tree level, (b. centre) $\mathbf{q}$-influence vector over a branch, (c. right) Change of $\Psi$-scores by increase of $\lambda_{leaf}$, \textbf{Bottom:} Change of $\Psi$-scores in a scale-free graph when (d. left) $\lambda_{deg=20}$ increases, (e. centre) $\mu_{deg=20}$ decreases, (f. right) E-R graph tragedy of commons.}
\label{perf fig}
\end{figure*}	

%%%%%%%%%%%%%% Analyse des plots %%%%%%%%%%%%%%%%%%%%%%%%%%%%

\section{Numerical Evaluation}
\label{numanal}

\subsection{Influence of graph topology and posting rates}

In this section we evaluate the performance of our analytical model and the $\Psi$-score of influence for various types of graphs:
\begin{itemize}
\item \textbf{Binary Tree:} We use a perfect undirected binary tree of depth $9$, which includes $N=1023$ nodes. Specifically, the $0$-node is the root, and there are $512$ leaves at level-$9$.
\item \textbf{Scale-free:} We build an undirected scale-free network of $N=50,000$ nodes with power-law degree distribution  exponent $2.5$, using a configuration model.
\item \textbf{Erd{\"o}s-R\'enyi (E-R):} For the undirected random graph with binomial degree distribution, we choose again $N=50,000$ nodes with mean degree $3$ neighbours per node. 
\end{itemize}
The evaluation for these graphs is illustrated in Fig.~\ref{perf fig}. We start with the binary tree. 

\textit{Node position:} For the binary tree, we evaluate the $\Psi$-score on the nodes of each level $\left\{0,1,\ldots,9\right\}$, where level-$0$ is the root level. Regarding user activity, we apply here the homogeneous case where all nodes have the same $(\lambda,\ \mu)$ pair. We evaluate for two activity choices $(0.25,\ 1)$ and $(1,\ 0.25)$ and plot the score-per-level at Fig.~\ref{tree_pos}. We observe that each level has a different score. The highest score is attributed to the parents of the leaf nodes (level-$8$), whereas the smallest score to the leaf nodes (level-$9$). The root node also gets a lower score, being a boundary one. Note that the $\Psi$-score will coincide with PageRank in this homogeneous case.

\textit{$\mathbf{q}$-influence:} Our model goes beyond PageRank even in the homogeneous activity case, to further explain in detail how each node's influence is distributed among the nodes of the network, through the vector $\mathbf{q}_i$. In Fig.~\ref{tree_q} we plot the influence $\mathbf{q}$-vector for $(\lambda,\mu)= (0.25,1)$ and three types of nodes $i=\left\{\text{leaf (level-$9$), leaf's parent (level-$8$), root (level-$0$)}\right\}$. Specifically, we select a single branch which spans the tree from root (level-$0$) to leaf (level-$9$), and plot the influence of $i$ on each level of this branch, i.e. $q_i^{(j)}$ for $j=0,\ldots,9$. There are several noteworthy observations. For the chosen activity values, self-influence $q_i^{(i)}$ is the highest for all three node types. Interestingly, the influence on direct neighbours is considerably higher than indirect ones, in all three cases. %Furthermore, for this specific choice of activity ratio $\lambda/\mu= 0.25/1$ the influence fades-out after $5$ levels away from the origin node. 
The leaf's parent collects the largest part of its $\Psi$-score from its influence on the leaf, because the latter does not have any other direct neighbours to be influenced by. On the other hand, the influence of the leaf to its parent is considerably less.

%%%%%%%%%%%%%% Plots de validation %%%%%%%%%%%%%%%%%%%%%%%%
\begin{figure*}[t!]
\centering
\hspace{-.5cm}
	\subfloat[Convergence for various arrival distributions]{\includegraphics[scale=0.40]{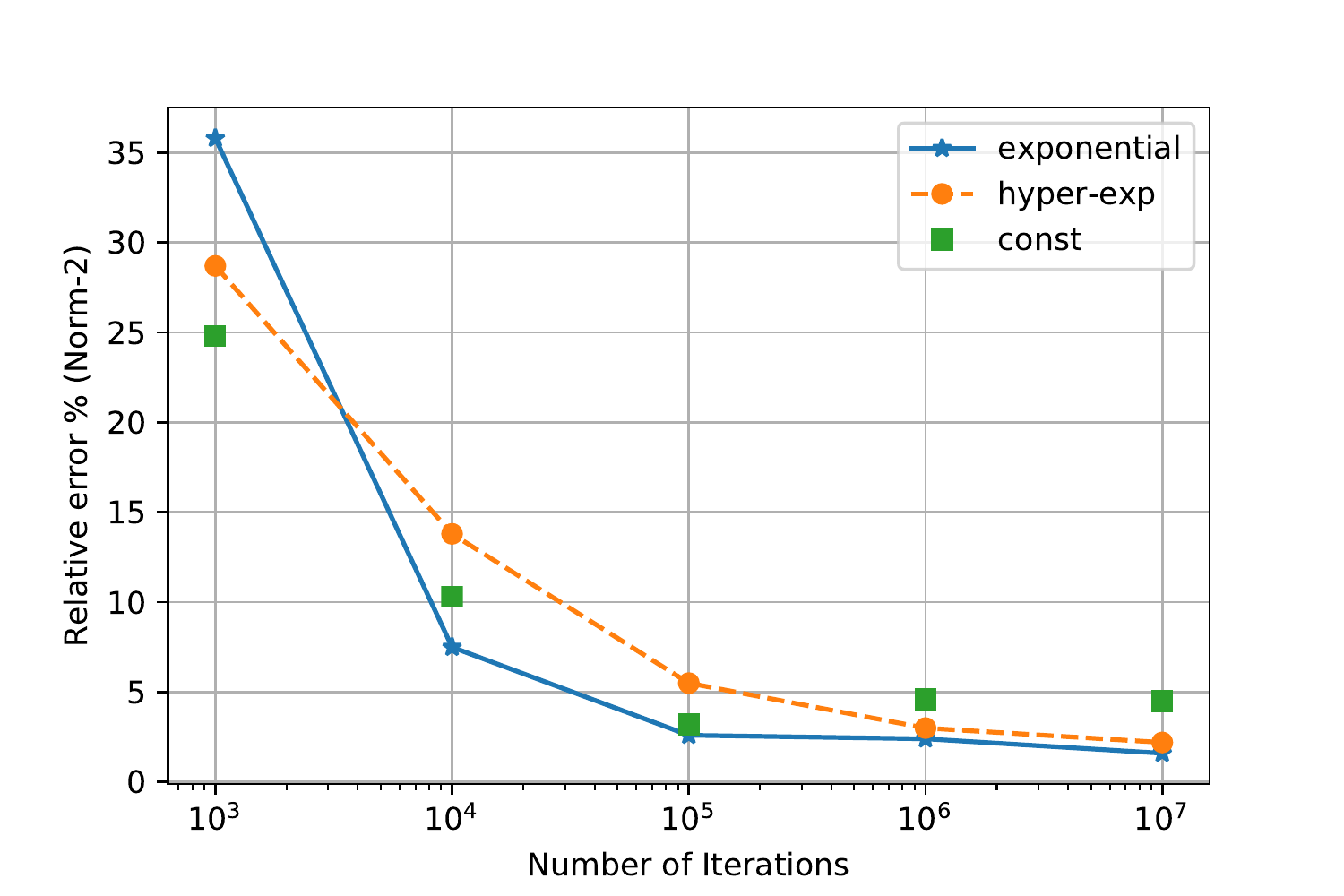}\label{robustproc}}
	\subfloat[Selection/Eviction policies]{\includegraphics[scale=0.35]{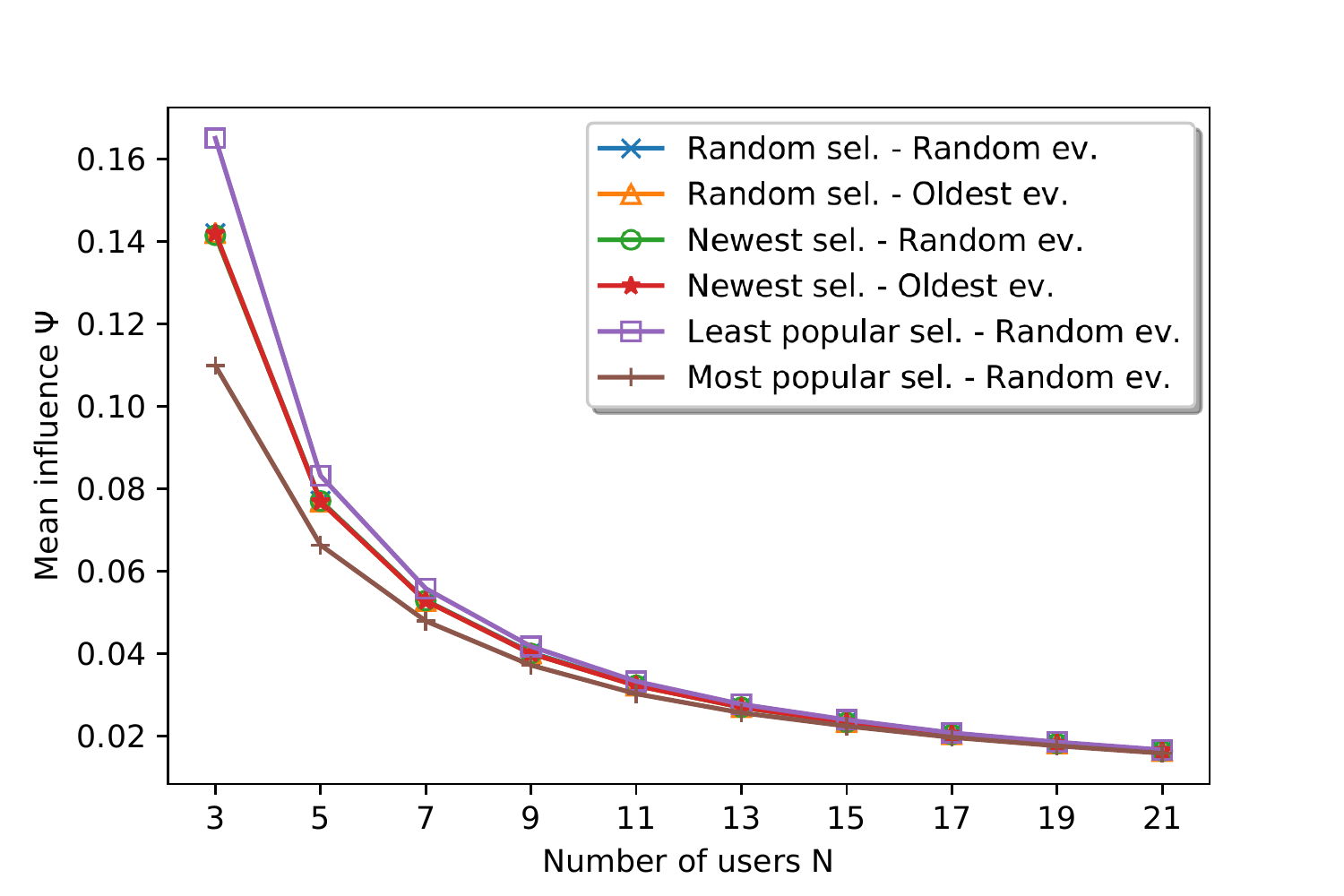}\label{robustselev}} 
	\subfloat[Influence per user]{\includegraphics[scale=0.35]{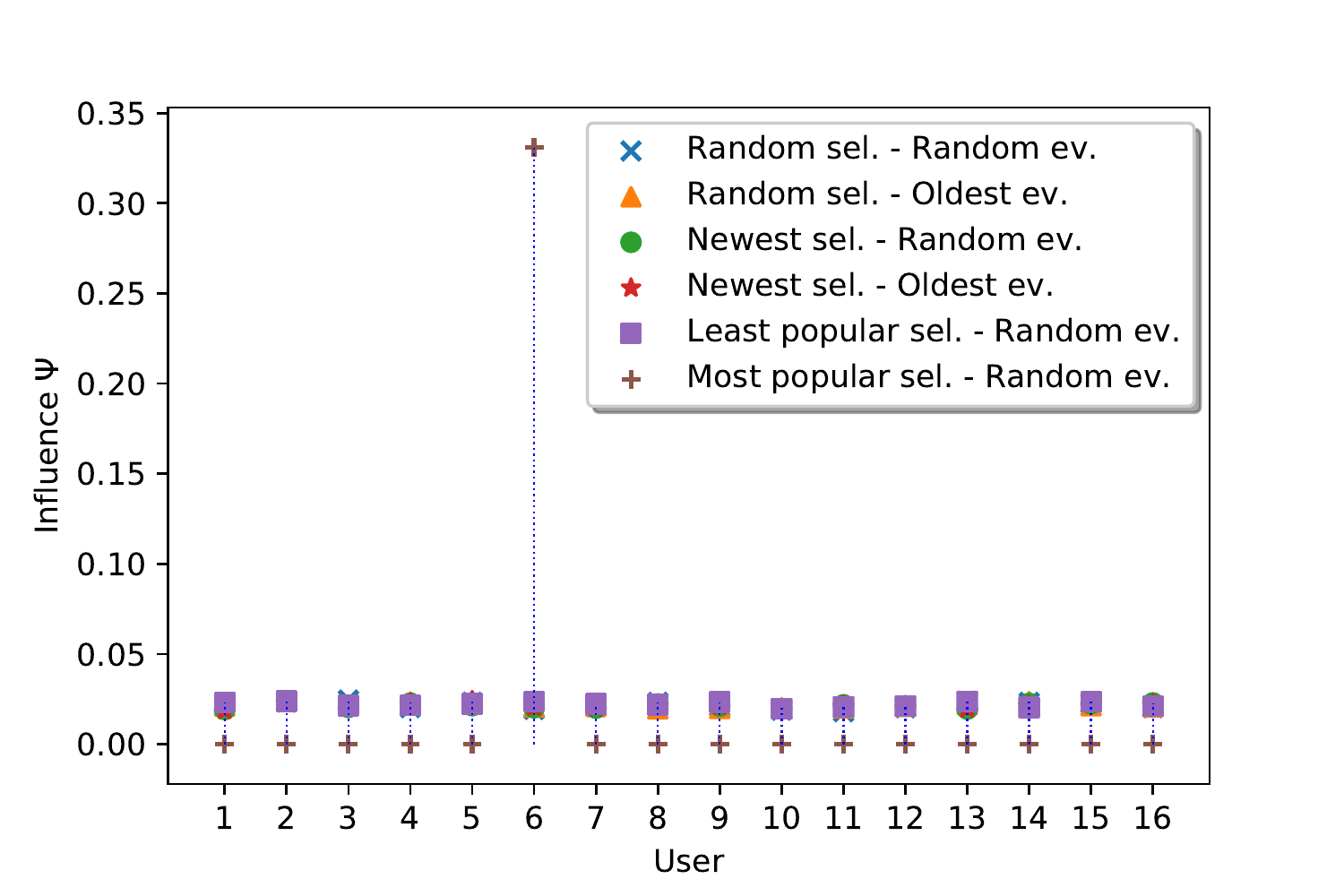}\label{robustuser}} 
\caption{(a)-(c): Sensitivity with respect to modeling assumptions.}
\label{valid fig}
\end{figure*}
%%%%%%%%%%%%%% %%%%%%%%%%% %%%%%%%%%%%%%%%%%%%%%%%%	

\textit{Increasing the leaf's posting rate:} 
Our model generalises PageRank, since it can calculate the $\Psi$-scores for nodes with asymmetric activity. We allow the $\lambda_{leaf}$ of a given leaf to increase between $[0.25,2.5]$, whereas all others keep their posting activity $\lambda=0.25$, and we calculate the new $\Psi$-scores of all nodes. In Fig.~\ref{tree_lam} we illustrate how the score evolves for the leaf node, its parent and the root node. We see that as $\lambda_{leaf}$ increases the $\Psi_{leaf}$ increases and the $\Psi_{leaf's\ parent}$ decreases. In fact, for $\lambda_{leaf}>1$, the ranks are inverted and the leaf node becomes more influential than its parent. The root's score does not change much, because of its distance from the leaf ($9$ levels away). As a conclusion the leaf has a naturally low score due to its position but can improve it
by increasing its posting rate.

\textit{Scale-free: }Next, we evaluate the $\Psi$-scores for the nodes of the scale-free graph. We first fix the same pair $(\lambda,\mu)=(0.25,1)$ for all nodes and select a specific target node with $degree=20$. For this node, we increase its posting rate $\lambda_{20}\in[0.25,2.5]$ and plot the change of its $\Psi$-score, as well as the score of some of its neighbours and non-neighbours in Fig.~\ref{sf_lam}. In this graph the degree plays an important role and for homogeneous $\lambda$ and $\mu$ the $\Psi$-scores are almost proportional to the degree, as can be expected from its relation with PageRank. But, as in the tree case, we observe that the score of the target node increases as $\lambda_{20}$ increases, whereas the score of its direct neighbours diminishes. The effect is more pronounced for a neighbour with degree $1$, since the target node is the only neighbour that can be influenced by this one; the effect is less strong but still visible for a neighbour of degree $4$. Other nodes of same degree $1$ and $4$ that are not direct neighbours of the target node do not exhibit a change in their score.

The importance of re-posting rate $\mu$ can be visualised in Fig.~\ref{sf_mu} where we fix $(\lambda,\mu)=(0.25,1)$ for all nodes, but decrease $\mu_{20}$ of the same target node with degree $20$ as before, while keeping $\lambda_{20}=\lambda$ fixed. In this scenario, we observe that as the target's re-posting decreases, its score increases, while the scores of its direct neighbours with degree $1$ and $4$ decrease as expected. The reduction is more pronounced for the degree $1$ direct neighbour. Again, the scores of nodes further away are less sensitive to changes in $\mu_{20}$.

\textit{A tragedy of the commons:} We have seen in the tree and scale-free graph that a node can improve its score if it posts with a larger $\lambda_i$. The same phenomenon can be observed in the Erd\"os-R\'enyi case. We will study now what happens when more and more nodes adopt such strategy and increase their posting rates in this example of E-R graph. For Fig.~\ref{er_tragedy} we consider a random graph where initially all nodes have $\lambda<\mu$, specifically $(\lambda,\mu)=(0.25,1)$. These nodes are marked with orange colour and we call them ``listeners''. Gradually more and more nodes adopt the strategy to increase their posting rate and decrease their re-posting rate to $(1,0.25)$ in order to improve their individual score. These nodes are marked with blue and we call them ``influencers''. The plot illustrates how the average score of listeners and influencers changes as the number of influencers gradually increases. For $5,000$ influencers and $45,000$ listeners, the influencers' average score is much higher than that of the listeners and their strategy to change activity bears fruit. But, as the number of influencers increases, both types get a smaller score because there are less nodes in the graph willing to be influenced. As soon as the influencers become the majority, the score of both types becomes less than the initial average score when everyone was a listener. At the extreme case where everyone is an influencer, the average $\Psi$-score is three times less than the other extreme where everyone is a listener. We observe here a typical case of the tragedy of the commons.

\subsection{Robustness} \label{robust}

We further evaluate the robustness of the model with respect to the modelling assumptions. For this purpose, we first show the convergence time of our simulator to the theoretical value. We then modify our simulator to take into account inter-arrival distributions alternative to Poisson, as well as selection and eviction policies other than random.

\paragraph{Convergence} We evaluate the convergence speed of the simulator to the solution of the balance equations by increasing the number of iterations $T\_iter$ and plotting the relative error; this is the ratio of the 2-norm of the difference between simulated and analytical values, divided by the 2-norm of the analytical values. We choose for the experiment a ring graph of $N=8$ with random $\lambda_i\in[0,1]$, $\mu_i\in[0,1]$ per user and random association of $4$ followers on average per node. The convergence till a relative error of $1.5\%$ is shown in Fig.~\ref{robustproc}.

\paragraph{Alternative inter-arrival times}
In Fig.~\ref{robustproc}, we plot in addition to the exponential inter-arrival case, also the convergence of the simulator to the solution from the balance equations when applying alternatively the following two distributions for both posting and re-posting:
(i) hyper-exponential (with same mean but higher variance than Poisson) and (ii) deterministic (with same mean and zero variance). The figure shows that the hyper-exponential converges to the same solution as the Poisson albeit more slowly. It exhibits a $1.6\%$ relative error compared to the analytical solution for $T\_iter=10^{7}$. The convergence for fixed interval process is non-monotone with a relative error of $5\%$ for $T\_iter=10^{7}$. Intuitively, since the balance equations are based on the conservation law of posts, these will hold for any arrival distribution that guarantees convergence to some steady-state distribution $\mathbf{p}$ and $\mathbf{q}$.

\paragraph{Policies}
We use here a complete graph with a varying number $N$ of users, and set $(\lambda,\mu)=(10,5)$. We programmed our simulator (see \ref{simu}) to test alternative policies for user selection and post eviction, based on age and post-popularity. In (\ref{pol}) we showed that certain alternatives give the same balance equations as the basic ``Random/Random'' policy; specifically the ``newest selection'' policy, where a user always chooses to re-post the most recent post on his/her Newsfeed, and the ``FIFO (oldest) eviction'' where the new post enters the top of the list and pushes out the oldest one. Figure~\ref{robustselev} shows that indeed ``Random/Random'', ``Newest/Random'', ``Random/FIFO (oldest)'' and ``Newest/FIFO'' have the same average performance, as these curves coincide. Furthermore, we test the ``least popular (resp. most popular) selection'', where the user chooses to re-post from his/her Newsfeed the post with the maximum (resp. minimum) current global number of re-posts. These policies use extra information on re-post history. We observe in Fig.~\ref{robustselev} that with such choices, the absolute difference with a random policy becomes higher. The comparison is more pronounced when we observe the individual influences on a complete graph with $16$ users in Fig. \ref{robustuser}. Interestingly, all policies behave similarly, except the "most popular" selection, which tends to give all influence to a single user. We conclude that our model is robust regarding the choice of selection and eviction policies when only local information involved. For other cases further investigations should shed more light on this very interesting question.

%%%%%%%%%%%%%% Traces %%%%%%%%%%%%%%%%%%%%%%%%%%%%

\section{Numerical results from Real-world Traces}
\label{trace}
\textbf{Traces:} We evaluate our model and $\Psi$ metric using two real datasets, one from Twitter and the second from the Weibo social platform. The first trace found in Kaggle is referred to as $\texttt{Russian}$\footnote{Available at  \href{https://www.kaggle.com/borisch/russian-election-2018-twitter}{https://www.kaggle.com/borisch/russian-election-2018-twitter}}. It contains roughly 2 million (re-)tweets emitted from $180,000$ users during the Russian presidential elections of 2018. The second trace referred to as $\texttt{Weibo}$\footnote{Available at \href{https://aminer.org/influencelocality}{https://aminer.org/influencelocality} from the paper \cite{Zha13}.} comes from \cite{Zha13} and contains roughly 34 million messages exchanged over the Chinese microblogging platform Sina Weibo. For both traces, user IDs are anonymised. Posts are ordered in time and their content is removed. Each line contains: $[\mathrm{PostID,\ TimeStamp,\ UserID,\ RePostID}]$. Basic statistics for both datasets are summarised in \autoref{data_stats}. The related friendship graphs are both sparse with an average in-degree (\#Followers) of 5.70 for $\texttt{Russian}$ and 236.9 for $\texttt{Weibo}$ and their degree distribution is heterogeneous and close to a power law. Their statistics are summarised in \autoref{comparegraphs}.

\textbf{Methodology:} 
Before starting, we need to extract some information from the traces. For each available data trace the input vectors of user activity $\lambda_i,\ \mu_i,\ \forall i$ can be estimated as the sample means of each user's posting activity, $\hat{\lambda}_i,\ \hat{\mu}_i,\ \forall i$ i.e., the ratio of the number of posts or re-posts over the total trace duration; the social follower graph $\mathcal{G}$ if not provided should also be inferred from the available traces, as we will show later in the case of $\texttt{Russian}$.

For each trace, we want to show how well our model can evaluate user influence in the respective platform. 
First, we rank the users based on the model influence $\Psi^{\mathrm{model}}$, where we use the sparse version of the code described in Section \ref{algo} with input $(\hat{\lambda},\hat{\mu},\mathcal{G})$, and compare this list with the ranking based on user empirical influence, as derived from the emulator $\Psi^{\mathrm{emu}}$ in Section \ref{emuli}. 
The \textit{empirical influence} ${q}^{emu}[i][j]$ in a trace is equal to the percentage of time that posts of origin $i$ occupy the first position on the Wall of user $j$, assuming a FIFO principle and $K=1$ Wall size.

As a second step we compare the model ranking, with the ranking based on alternative influence measures: the user (i) number of followers (in-degree), (ii) post activity $\hat{\lambda}$, (iii) PageRank \cite{pagerank98}. 

To compare two ranking lists between each other we use two types of plots: A 2D scatter plot, where each point corresponds to a user and is the tuple of his/her predicted rank based on the $\Psi^{\mathrm{emu}}$ (x-axis) and the $\Psi^{\mathrm{model}}$ (y-axis); in such plots we also visualise the distance of the rankings from the line $x=y$, which describes the ideal perfect match of ranks. The second type of plot illustrates a metric similar to the Jaccard index to compare the two rankings, called ``Common users proportion". More precisely, if $\{u_1, \ldots, u_X\}$ are the top-$X$ $\mathrm{UserID}$s according to the emulator list, whereas $\{v_1, \ldots, v_X\}$ are the top-$X$ $\mathrm{UserID}$s for the model list, we define the proportion of common users at depth $X$ of the emulator list by

	\[ \mathcal{C}_X = \frac{|\{u_1, \ldots, u_X\} \cap \{v_1, \ldots, v_X\}|}{X}. \]
Note that the quantity $\mathcal{C}_X$ converges to $1$ as $X$ grows to $N$, because the two full lists contain the same set of users. But for some given $X<N$ (e.g. top-$10$), the curve shows how well the model manages to rank users in relation to the emulator in the top-$X$ positions.

%%% PLOTS RUSSIAN %%%
\begin{figure}[t!]
	\centering
	{\includegraphics[scale=0.4]{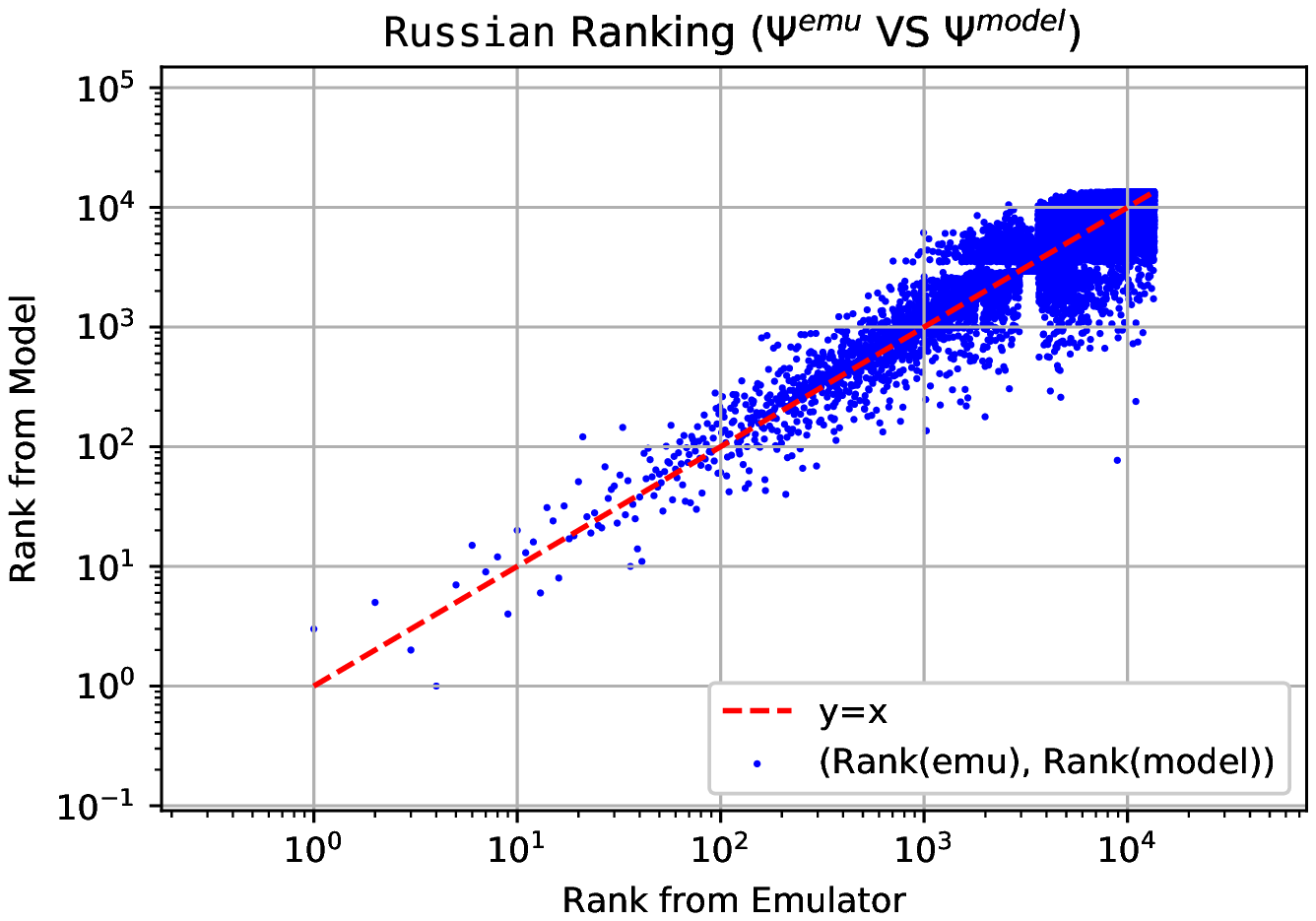}\label{russian_rank_scatter_star}}		
	{\includegraphics[scale=0.4]{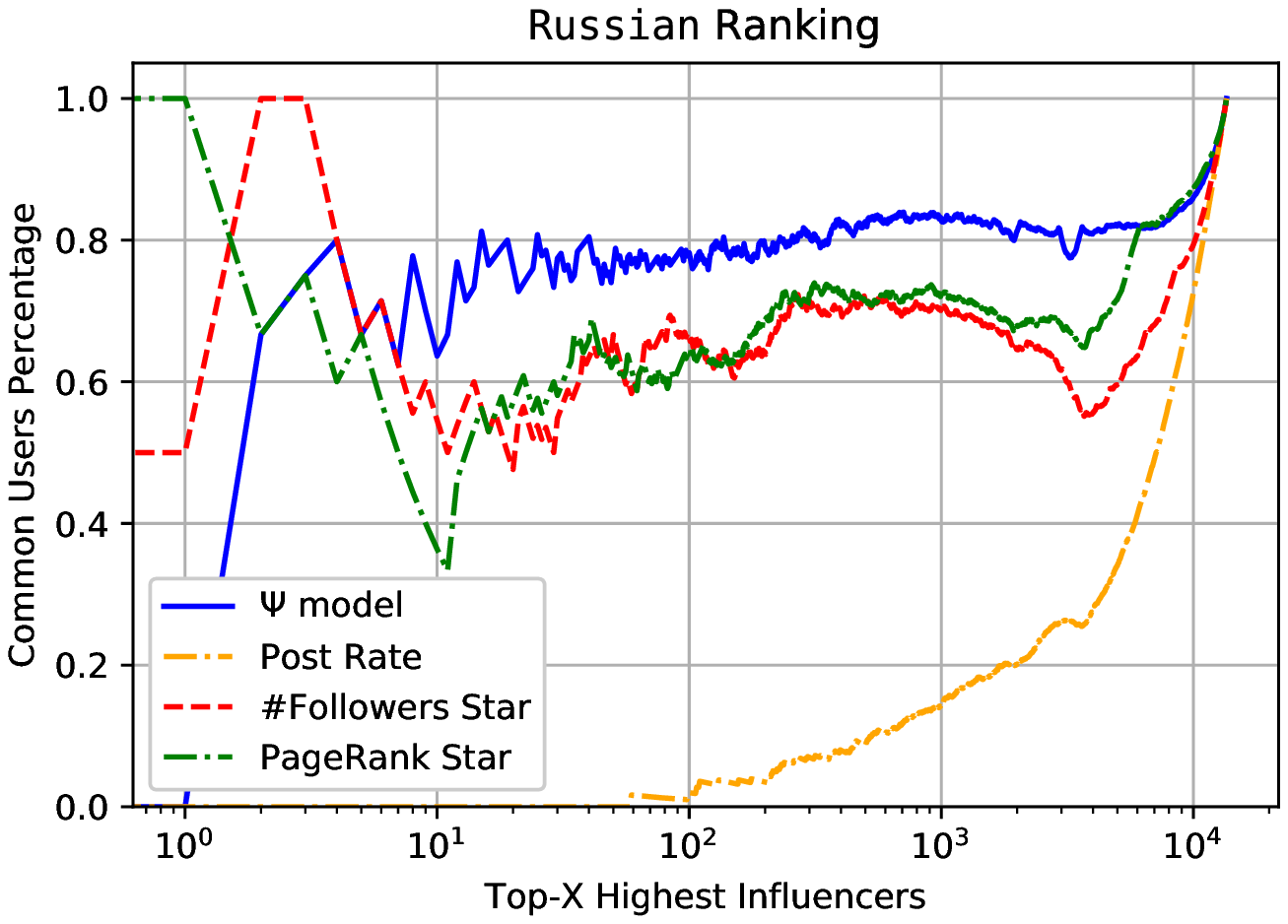}\label{russian_common_users}}
	\caption{User ranking comparison for $\texttt{Russian}$. (a) top: Scatter plot between Rank$(\Psi^{\mathrm{model}})$ and Rank$(\Psi^{\mathrm{emu}})$ -- points on the left rank higher ($1st$, $2nd$, etc.), (b) bottom: Common users proportion with reference to Rank$(\Psi^{\mathrm{emu}})$.
	}
	\label{plots russian}
\end{figure}

%%% DATA STATS TABLE %%%
\begin{table}[t!]
	\centering
	\caption{Basic statistics on both datasets.}
	\def\arraystretch{1.2}
	\begin{tabular}{l|c||c|}
		\cline{2-3}
		& \textbf{Russian} & \textbf{Weibo} \\ \hline
		\multicolumn{1}{|l|}{\textbf{Time window}} & 57 days & 1 216 days \\ \hline
		\multicolumn{1}{|l|}{\textbf{\# users}} & 181 621 & 1 340 816 \\ \hline
		\multicolumn{1}{|l|}{\textbf{\# posts}} & 674 292 & 232 978 \\ \hline
		\multicolumn{1}{|l|}{\textbf{\# reposts}} & 1 271 073 & 33 307 189 \\ \hline
		\multicolumn{1}{|l|}{\textbf{Mean \#posts/user}} & 3.71 & 0.17 \\ \hline
		\multicolumn{1}{|l|}{\textbf{Mean \#reposts/user}} & 7.00 & 24.84 \\ \hline
		\multicolumn{1}{|l|}{\textbf{Max \#posts}} & 4 834 & 3 718 \\ \hline
		\multicolumn{1}{|l|}{\textbf{Max \#reposts}} & 2 811 & 1 032 \\ \hline
		\multicolumn{1}{|l|}{\textbf{\% users with \#posts $> 0$}} & 54.45 & 03.55 \\ \hline
		\multicolumn{1}{|l|}{\textbf{\% users with \#reposts $> 0$}} & 63.60 & 99.54 \\ \hline
	\end{tabular}
	\label{data_stats}
\end{table}

%%% TABLE GRAPH STATS RUSSIAN AND WEIBO %%%
\begin{table}[t!]
	\centering
	\caption{Statistics of input social graphs.} 
		\begin{tabular}{c|c||c|}
			\cline{2-3}
			\multicolumn{1}{l|}{} & \textbf{Russian} (Star) & \textbf{Weibo} (Real)  \\ \hline							\multicolumn{1}{|c|}{\textbf{\#nodes}} & 181 621 & 1 340 816 \\ \hline 
			\multicolumn{1}{|c|}{\textbf{\#edges}} & 517 421 & 291 761 716 \\ \hline 
			\multicolumn{1}{|c|}{\textbf{mean \#followers (in-degree)}} & 5.70 & 236.90 \\ \hline 
			\multicolumn{1}{|c|}{\textbf{max \#followers (in-degree)}} & 7 868 & 431 385 \\ \hline  
			\multicolumn{1}{|c|}{\textbf{max \#leaders (out-degree)}} & 389 & 8 107 \\ \hline 
		\end{tabular}
	\label{comparegraphs}
\end{table}

%%% TABLE TOP INFLUENCERS RUSSIAN %%%
\begin{table}[t!]
	\centering
	\caption{Top-10 influencers in $\texttt{Russian}$ as returned by the emulator and compared with the model (star graph).}
	\begin{tabular}{|l|c|c|c|c|c|c|}
		\hline
		\textbf{User} & $\boldsymbol\Psi^{\textbf{emu}}$& $\boldsymbol\Psi^{\textbf{model}}$ & $\textbf{Rank}$ & $\textbf{Rank}$ & \textbf{\#Follow} & $\boldsymbol{\lambda}[s^{-1}]$ \\
		ID \# & $10^{-3}$ & $10^{-3}$ & emu & model & Star & $10^{-7}$ \\
		 \hline
		\textbf{20905367} & 12.02 & 10.23 & 1 & 3 & 6 676 & 42.9 \\ \hline
		\textbf{82299300} & 11.29 & 7.03 & 2 & 5 & 7 833 & 96.0 \\ \hline
		\textbf{494076761} & 7.35 & 13.80 & 3 & 2 & 6 963 & 439.3 \\ \hline
		\textbf{615422017} & 5.33 & 13.82 & 4 & 1 & 5 474 & 639.5 \\ \hline
		\textbf{174953869} & 5.13 & 5.81 & 5 & 7 & 1 742 & 118.5 \\ \hline
		\textbf{711363811} & 4.79 & 4.12 & 6 & 15 & 1 309 & 8.2 \\ \hline
		\textbf{36309919} & 4.44 & 5.23 & 7 & 9 & 4 516 & 118.5 \\ \hline
		\textbf{1867848452} & 4.15 & 4.60 & 8 & 12 & 1 235 & 6.1 \\ \hline
		\textbf{34200559} & 3.68 & 7.96 & 9 & 4 & 3 571 & 982.8 \\ \hline
		\textbf{50597428} & 3.36 & 3.64 & 10 & 20 & 1 156 & 8.2 \\ \hline
	\end{tabular}
	\label{russian_psi_table}
\end{table}

\subsection{First Dataset --- $\texttt{Russian}$}

To apply our sparse algorithm (see \ref{algo}) we estimate user activity by the sample means $(\hat{\lambda},\ \hat{\mu})$, i.e. the number of posts and reposts per user in the trace divided by the total time window ($57$ days). The user graph needed as input is not directly available for $\texttt{Russian}$ and we have to find a way to recover it. For this we use the following heuristic:
\begin{description}
	\item[\textbf{Star} $\mathcal{G}$] $\ \ $ User $i$ is considered to be a follower of user $j$ iff $i$ has reposted a post of origin $j$ at least once. 
\end{description}
With \textit{Star} we make the simplifying assumption that a user only reposts content created by his/her direct leaders. This approach short-circuits the diffusion paths of posts as it links directly original authors to all the re-posters, drawing a network with star-shaped communities. The \textit{Star} graph statistics are given in \autoref{comparegraphs}. We chose to follow this approach because the data-trace does not contain extra information over the IDs of ``relay''-users.
In such traces we can only know and store the original author and the total set of users each post reached, but not the paths. Note here that inferring graphs from data is a subject of active research and alternative methods can be found in the recent literature \cite{NewmanRich}. For the time being we will use the Star-network as input; in the \texttt{Weibo} section, we will benefit from a real known graph.

The evaluation of the $\texttt{Russian}$ model against the emulator is provided in \autoref{russian_psi_table}, and in \autoref{plots russian}. The top plot in \autoref{plots russian} shows the scatter plot of ranking based on $\Psi^{\mathrm{model}}$ and $\Psi^{\mathrm{emu}}$; note here that the points at the left correspond to higher influence rank ($1st$, $2nd$,...) compared to points at the right, which are less significant. We observe a very good fit between the two rankings in the entire domain, and even for the top ranks  ($1st$, $2nd$,...) at the left part of the plot, which are of practical interest, e.g. for a company targeting high-influencers. At the bottom plot we illustrate four curves: each of them plots the Common users proportion between the user ranking from the emulator and the ranking from (i) the model influence $\Psi^{\mathrm{model}}$ using the Star graph, (ii) user post rate, (iii) user number of followers from the model, and (iv) user PageRank from the Star graph with damping factor $\beta=0.85$. We observe that our model (i) explains much better than the other metrics the ranking by the emulator, and is able to find $80\%$ of emulator top-$X$ users, in the largest part of the $X$ range. The two curves (iii) and (iv) which describe only the graph structure perform much lower. Finally, in this specific example, the curve (ii) about user posting rate has no considerable importance in explaining influence, as such ranking seems unrelated to the emulator.

Precise results for the top-10 $\Psi^{\mathrm{emu}}$ ranked users are shown in \autoref{russian_psi_table}. We observe a very good fit between actual values of $\Psi^{\mathrm{model}}$ and $\Psi^{\mathrm{emu}}$. Specifically, the model manages to find 7-out-of-10 influencers in the emulator top-10 list. An important observation comes from the two last columns of the table, namely the user number of followers and the user posting rate. We see that neither the number of followers nor the posting frequency follows the ranked order of the top influencers, in support of our observations already in \autoref{plots russian}. By inspection, the user with maximum \#followers ($7833$ followers in Star = $7833$ users who shared posts originating from him/her) is ranked $2^{nd}$ by the emulator and $5^{th}$ by the model, whereas the user with maximum posting rate is ranked $9^{th}$ by the emulator and $4^{th}$ by the model. Impressively, we find in the list a top-influencer (ranked $8^{th}$ by the emulator) with low posting rate (e.g. $6.13\cdot 10^{-7}$ [posts/sec] $\approx0.00221$ [posts/day]) and relatively low \#followers ($1235$ ``followers'' in Star). Hence, neither the \#followers as a measure of importance in the social graph, nor the posting rate as an activity measure are alone sufficient to rank users by influence. Our $\Psi$-score mixes user activity with graph position.

For \texttt{Russian}, an important part of the user influence is already present in the Star-graph, due to the way we chose to draw it (all users who shared a post are assumed as followers of its original author). The correlation between the influence score $\Psi^{\mathrm{model}}$ and \textit{Star} in-degree (\#followers), is very high, equal to $0.82$, whereas the correlation between $\Psi^{\mathrm{model}}$ and posting activity $\lambda$ is only $0.11$. For the above reasons, we study the second dataset ( \texttt{Weibo}), accompanied by its true social graph.

\subsection{Second Dataset --- $\texttt{Weibo}$}
\label{weibo}

For the second dataset ($\texttt{Weibo}$) we have access to the underlying friendship graph i.e., who follows whom. The graph statistics are given in \autoref{comparegraphs}\footnote{The friendship graph contains users that do not appear in the trace; since no further information is available over their activity we ignore them.}. We use this \textit{Real} graph to compute values of influence from our model $\Psi^{\mathrm{model}}$, whereas the $\Psi^{\mathrm{emu}}$ is derived directly from the trace. The \texttt{Weibo} trace is special, because of the way it was collected to serve the study of cascades (see \cite{Zha13}); the authors isolated and kept in the dataset only certain (approx. $200K$) microblog episodes, which were massively re-posted. Specifically, as can be seen from \autoref{data_stats} there are $10\times$ more users than the \texttt{Russian}, but only $3.5\%$ of users post, in comparison to $99.5\%$ who re-post. The number of original posts is very small compared to much larger ($100\times$) number of re-posts, meaning that the trace has a small user set of potentially very large influence. Here, we expect user activity to play an important role in determining influence, not just graph structure. On the other hand, the Real graph contains $291$ million (M) edges (see \autoref{comparegraphs}), most of which are never observed to be active in the available trace for post forwarding. In fact the trace contains $\approx 33$M re-posts. Even if each repost passed through a different edge from the Real graph, there would be $291-33=258$M edges unused. Let us see how our model behaves in this particular case.

\begin{figure*}[t!]
	\centering
	{\includegraphics[scale=0.4]{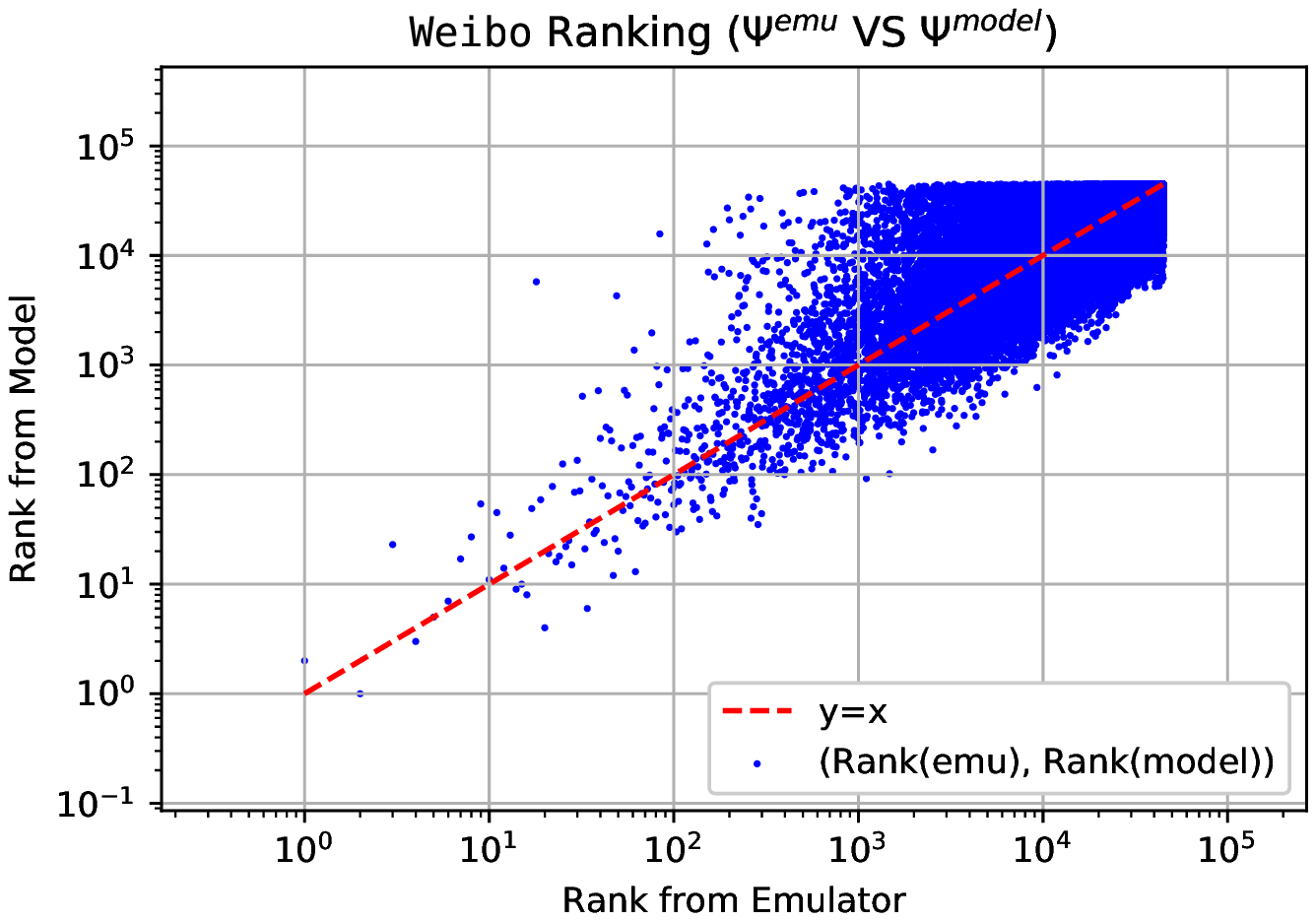}\label{weibo_realrank}}
	{\includegraphics[scale=0.4]{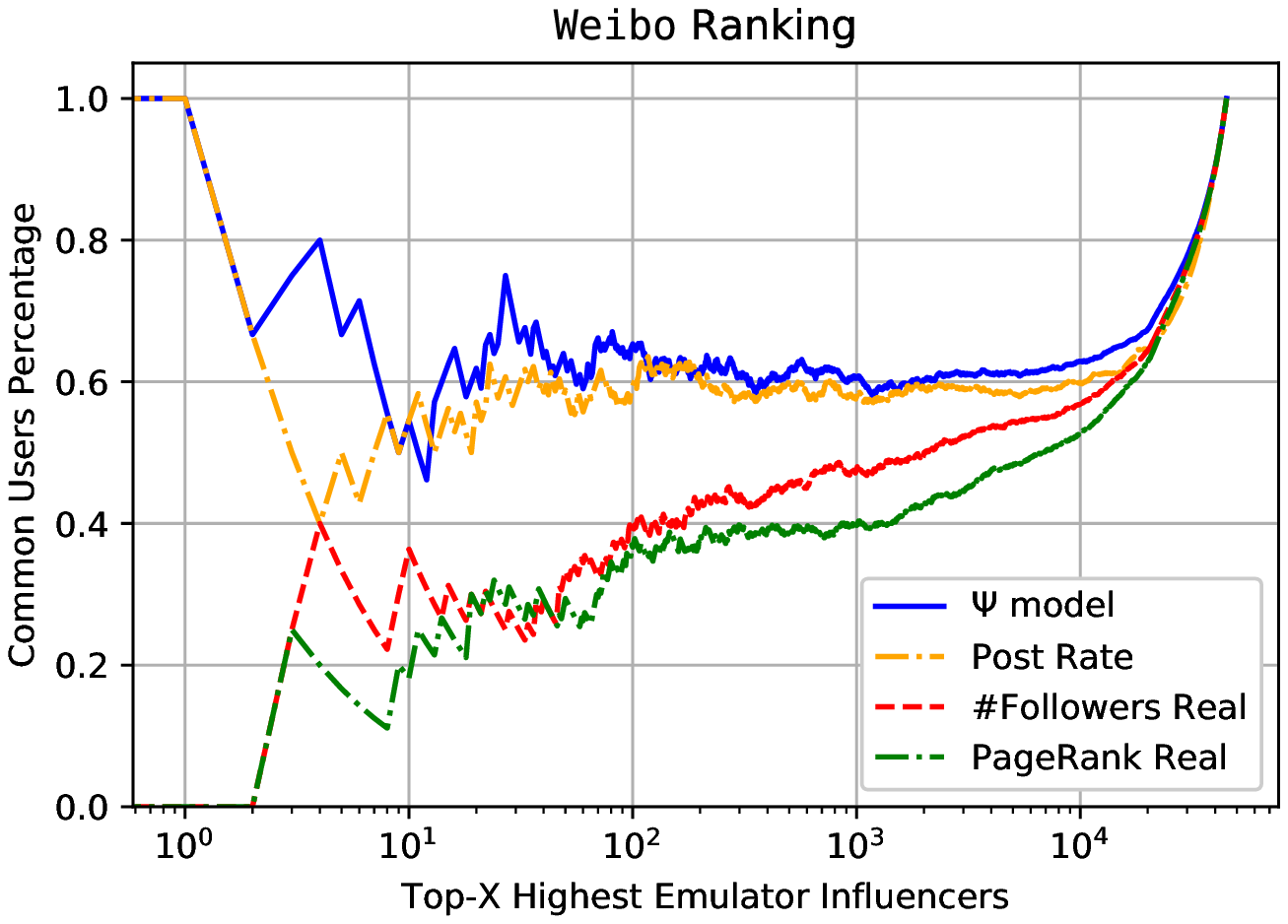}\label{weibo_commonALL}}
	{\includegraphics[scale=0.4]{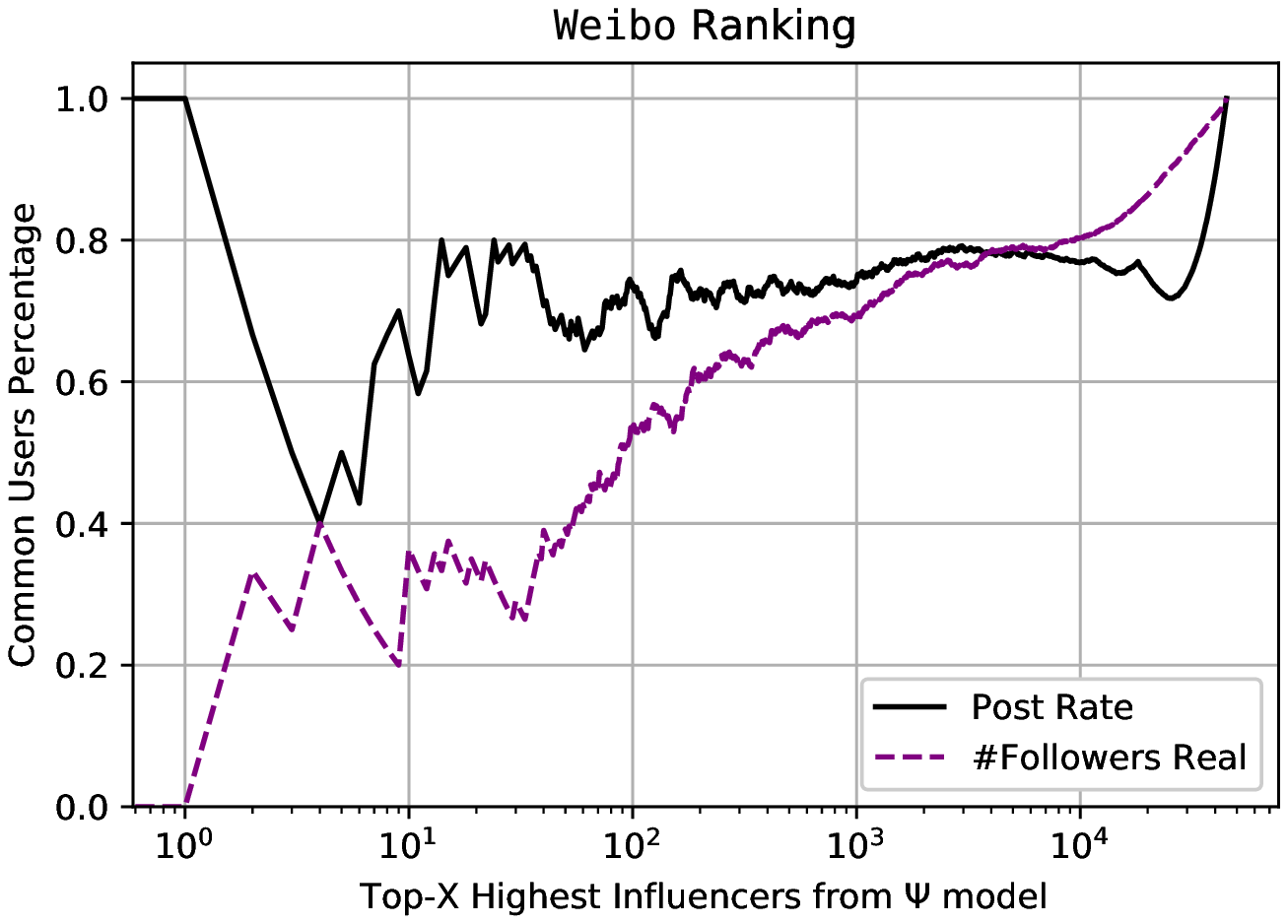}\label{weibo_explain}}
	\caption{User ranking comparison for $\texttt{Weibo}$. (a) left: Scatter plot  between Rank$(\Psi^{\mathrm{model}})$ and Rank$(\Psi^{\mathrm{emu}})$ -- points on the left rank higher ($1st$, $2nd$, etc.), (b) centre: Common users proportion with reference to Rank$(\Psi^{\mathrm{emu}})$, (c) right: Common users proportion with reference to Rank$(\Psi^{\mathrm{model}})$.
	}
	\label{plots weibo}
\end{figure*}

User ranking for \texttt{Weibo} using the model and the emulator are shown in \autoref{plots weibo} and \autoref{Weibo-top10}. Specifically, \autoref{plots weibo} (left) presents the scatter plot for ranking by $\Psi^{\mathrm{model}}$ and $\Psi^{\mathrm{emu}}$. The fit for high influencers is good, but worsens as we move to the right in the low influence ranks, but the points are always centered around the line $x=y$. This behaviour is partly due to the specific trace and partly due to numerical issues; the massive size and density of the \texttt{Weibo} graph and the asymmetry in activity slowed-down computation of $\Psi^{\mathrm{model}}$, when using our sparse algorithm for the model, and we had to trade-off accuracy for run-time. 

\autoref{plots weibo} (centre) shows the Common users proportion metric between the rank list from $\Psi^{\mathrm{emu}}$ and the ranking from: (i) \#followers, (ii) PageRank (with $\beta=0.85$), (iii) post rate, and (iv) the model $\Psi^{\mathrm{model}}$.  The performance of $\Psi^{\mathrm{model}}$ is again the best and it can explain around $60\%$ of the user top-$X$ ordering by $\Psi^{\mathrm{emu}}$. The reason for the lower performance compared to \texttt{Russian} \textit{Star} is that, in the Real graph most of the edges do not participate in the post diffusion described in the \texttt{Weibo} trace. Our model in this paper does not include contextual preferences towards users or topics, but rather gives equal probability to all posts visible in the Newsfeed to be re-posted. We believe that a pre-processing of the real graph to keep only active relationships (edges) could significantly improve the model performance. Even in this unfavourable situation, however, we observe once again that our model using the Real graph outperforms the other three measures in explaining the top-X influencers found in the emulator ranking list. Interestingly, the graph-related measures of in-degree (\#followers) and PageRank behave very badly, whereas the post rate explains much better the empirical influence; this is to be expected because only cascades of posts from specific origins are kept in the trace. The $\Psi$ metric from our model combines both graph position and user activity to give a better estimation of social influence. Finally, \autoref{plots weibo} (right) compares ranking by the model with ranking by the post rate and the in-degree. Although $\Psi^{\mathrm{model}}$ and post rate performance seem very close in the (centre) plot, less than $80\%$ of the $\Psi^{\mathrm{model}}$ ranking can be explained just by the post rate. This means that the $\Psi^{\mathrm{model}}$ cannot be replaced simply by the post rate - even in this special case of dataset, because it contains different information over user influence.

Finally, referring to \autoref{Weibo-top10}, we see that the model with Real graph input finds 5-out-of-10 top influencers common with the emulator. From the \#followers and activity $\lambda$ columns, we verify again that the influence $\Psi$-score cannot be explained by \#followers (in-degree) or activity only, but rather by an appropriate combination of the two, summarised in $\Psi^{\mathrm{model}}$.

%%% PLOTS WEIBO %%%

%%% TABLE TOP INFLUENCERS WEIBO %%%
	\begin{table}[!t]
	\centering
	\caption{Top-10 influencers in $\texttt{Weibo}$ as returned by the emulator and compared with the model (Real graph).}
		\begin{tabular}{|l|c|c|c|c|c|c|}
			\hline
			\textbf{User} & $\boldsymbol\Psi^{\textbf{emu}}$ & $\boldsymbol\Psi^{\textbf{model}}$ & $\textbf{Rank}$ & $\textbf{Rank}$ & $\textbf{Follow}$ & $\boldsymbol{\lambda}[s^{-1}]$ \\ 
			ID \# & $10^{-3}$  & $10^{-3}$ & $emu$ & $\textit{model}$ & $\textit{Real}$  & $10^{-6}$ \\ \hline
			\textbf{519514} & 37.08 & 63.88  & 1 & 2  & 459 & 31.7 \\ \hline
			\textbf{490872} & 24.68 & 93.99  & 2 & 1  & 595 & 35.4 \\ \hline
			\textbf{1004172} & 13.30 & 7.05  & 3 & 23  & 520 & 2.0 \\ \hline
			\textbf{482551} & 11.53 & 47.56  & 4 & 3 & 1247 & 11.0 \\ \hline
			\textbf{110361} & 7.07 & 13.19  & 5 & 5 & 288 & 10.3 \\ \hline
			\textbf{244531} & 7.05 & 12.33  & 6 & 7 & 312 & 10.4 \\ \hline
			\textbf{296675} & 6.77 & 8.30  & 7 & 17 & 347 & 8.7 \\ \hline
			\textbf{980392} & 6.70 & 5.94  & 8 & 27 & 230 & 6.2 \\ \hline
			\textbf{153610} & 6.22 & 2.93  & 9 & 54 & 81 & 3.6 \\ \hline
			\textbf{821785} & 6.21 & 11.32  & 10 & 11 & 1084 & 2.7 \\ \hline
		\end{tabular}
		\label{Weibo-top10}
	\end{table}

%%%%%%%%%%%%%% Conclusions %%%%%%%%%%%%%%%%%%%%%%%%%%%%

\section{Conclusions}
\label{conclu}
In this work we have introduced an original Markovian model that analyzes the diffusion of posts in a generic social platform, and quantifies the influence of a given user over any other. By resolving it we have derived closed-form expressions for metrics of influence, which allow to rank users based on the novel $\Psi$-score; the latter summarises the combined effect of user position in the graph with user activity. These results constitute a novel powerful toolbox that can be further exploited to understand and design social platforms. To highlight the importance of these results we have implemented a sparse version of the solution algorithm in \cite{code} and have applied it to massive data traces from real OSPs (Twitter and Weibo). The model-based $\Psi$-ranking is verified to correspond well with the empirical influence as read from the traces, a fact which validates our model for real world applications. As a consequence, we believe that the model can serve as \textit{a test and prediction tool} for many social platform types and user activity scenarios, when no trace is available. It is flexible enough to further include extra OSP features (post-filtering, ``likes'', etc), or the balance equations could be modified to consider user preference towards posts from specific leaders or origins. The $\Psi-$score, itself, is shown more suitable to rank user influence compared to standard centrality metrics (\#followers, PageRank, user activity) and cannot be substituted by any of them. Hence, it enriches the literature with a novel important measure that can combine user position in the graph with user activity to adequately rank user influence inside a platform.

%% use section* for acknowledgment
\section*{Acknowledgment}
The authors would like to thank the anonymous reviewers for their insightful comments that helped improve the work.

%%%%%%%%%%%%%% BIBLIOGRAPHY %%%%%%%%%%%%%%%%%%%%%%%%%%%%

\bibliographystyle{alpha}

%%%%%%%%%%%%%% PROOF EXACTNESS %%%%%%%%%%%%%%%%%%%%%%%%%%%%

\appendix

\textbf{[A. Non-reversibility example]} Consider $N=3$ users $\left\{a,b,c\right\}$, each of whom is leader of the other two. Suppose also $M=1$ for the Newsfeed and $K=1$ for the Wall. In this simple case, the state description is simplified. Let the initial state of all Newsfeeds and Walls contain a post from $a$, so we write $news=(a,a,a)$, and $walls=(a,a,a)$ to summarise the $3$-user system state. Here, $news$ and $walls$ just show the current posts on the three Newsfeeds and the three Walls simultaneously, ordered user $a$ as 1st entry, user $b$ as 2nd and user $c$ as 3rd. Then with rate $\lambda^{(b)}$ user $b$ posts, so the posts on the $news=(b,a,b)$ and on the $walls=(a,b,a)$. With rate $\mu^{(c)}$ user $c$ reposts so $news=(b,b,b)$ and $walls=(a,b,b)$. With rate $\lambda^{(a)}$ we get $news=(b,a,a)$ and $walls=(a,b,b)$. With rate $\mu^{(b)}$ we get $news=(a,a,a)$ and $walls=(a,a,b)$. Finally, with rate $\mu^{(c)}$ we return to the initial state and the cycle is complete. However, the rate to traverse the cycle in reverse order is $0$, since there is no way we can transition from the initial state, to the next state $news=(a,a,a)$ and $walls=(a,a,b)$, because a post from $b$ can appear on the $c$ Wall only by re-posting, which is impossible in this case.\\

\textbf{[B. Proof of Theorem \ref{exact}]} We will show that (\ref{Cpj}) is exact (i.e., holds without any approximation). The proof is based on the \textit{conservation law of posts in the Newsfeed}. The same methodology can be used to prove exactness for the other three equalities (\ref{Cpi}), (\ref{Cqj}), (\ref{Cqi}).

Let us observe the Newsfeed of user $n$ at the steady-state. This user has a set of leaders $\mathcal{L}^{(n)}$ with index $k=1,\ldots,L$. Each leader has a posting activity, described by a homogeneous Poisson process (HPP) $N_p^{(k)}$ with rate $\lambda^{(k)}$ and a re-posting activity described by a HPP $N_r^{(k)}$ with rate $\mu^{(k)}$ respectively. The superposition of all $2L$ processes is itself a HPP $N$ of rate $\sum_{k\in\mathcal{L}^{(n)}}(\lambda^{(k)}+\mu^{(k)})$. This is the process of total incoming posts in the Newsfeed of user $n$. 
By fixing the time origin at $t_0=0$ and a time interval $(0,T]$ we count $N((0,T])$ incoming posts
\begin{eqnarray}
\label{countT}
N((0,T]) & = & \sum_{k=1}^L \left(N_p^{(k)}((0,T]) + N_r^{(k)}((0,T]) \right). 
\end{eqnarray}
We can write for the re-posts of user $k$ (and similar for posts)
\begin{eqnarray}
N_r^{(k)}((0,T]) & = & \sum_{s}\mathbf{1}_{(0,T]}(t_s),
\end{eqnarray}
where user $k$ reposted at random time instants $0< t_1<\ldots<t_{S}\leq T$, and the event count $S=N_r^{(k)}((0,T])$. 

We focus on posts of origin $i$. The state of the Newsfeed $n$ related to this type of posts at $t\in(0,T]$ is denoted by $X_i^{(n)}(t)\leq M$ and is itself a continuous-time process. The states of the leaders are denoted by $X_i^{(k)}(t)\leq M$, $k=1,\ldots,L$. Furthermore, the arrival process of posts of origin $i$ through leader $k$ is the counting process $N_{r,i}^{(k)}$, for $k=1,\ldots,L$. It holds $N_{r,i}^{(k)}((0,T])\leq N_{r}^{(k)}((0,T])$.

$\bullet$ \textbf{Incoming posts of origin $i$ to Newsfeed $n$:} Each leader $k$ when re-posting will choose at random a post from his/her own Newsfeed. This is the \textit{random selection} assumption. We can model the choice of user $k$ to re-post content with label $i$ at time $t<T$ as a Bernoulli random variable 
\begin{eqnarray}
\label{Zin}
Z_i^{(k)}(t)  & = & \left\{ \begin{tabular}{l l}
$1$ & , with probability $\frac{X_i^{(k)}(t)}{M}$\\
$0$ & , with probability $1-\frac{X_i^{(k)}(t)}{M}$
\end{tabular}
\right..
\end{eqnarray}
This is itself a random process, which depends on the current state of the leader's Newsfeed $X_i^{(k)}(t)$. The counting process $N_{r,i}^{(k)}$ results from thinning the re-post process $N_{r}^{(k)}$ of user $k$ based on the random variable (\ref{Zin}), and we write
\begin{eqnarray}
N_{r,i}^{(k)}((0,T]) & = &  \sum_{s}\mathbf{1}_{(0,T]}(t_s)Z_i^{(k)}(t_s). 
\end{eqnarray}
The expected number of posts of origin $i$ through leader $k$ is 
\begin{eqnarray}
\mathbb{E}\left[N_{r,i}^{(k)}((0,T]) \right]= \mathbb{E}\left[ \sum_{s}\mathbf{1}_{(0,T]}(t_s)Z_i^{(k)}(t_s)\right] & = & \nonumber\\
\mathbb{E}\left[\sum_{s}  \mathbb{E}\left[\mathbf{1}_{(0,T]}(t_s)Z_i^{(k)}(t_s)\ \left| \ t_s,X_i^{(k)}(t_s)\right.\right]\right]& = &   \nonumber\\
\mathbb{E}\left[ \sum_{s} \mathbf{1}_{(0,T]}(t_s)\frac{X_i^{(k)}(t_s)}{M}\right]. & &\nonumber
\end{eqnarray}
By the \textit{smoothing formula} for HPP of rate $\mu^{(k)}$ \cite[Th.7.1.7]{Brem20},
\begin{eqnarray}
\label{expectk}
\mathbb{E}\left[N^{(k)}_{r,i}([0,T])\right] & = & \mathbb{E} \left[ \int_{0}^T\frac{X_i^{(k)}(t)}{M}\mu^{(k)}dt \right].
\end{eqnarray}

For the expected total incoming number of posts of origin $i$ in Newsfeed $n$, we consider re-posts of origin $i$ through the $L$ leaders, plus self-posts by user $i$, if user $i$ is a leader of $n$,
\begin{eqnarray}
\label{inT}
& &\hspace{-1.cm}\mathbb{E}\left[N_{in,i}((0,T])\right] = \nonumber\\
& & =\lambda^{(i)}T\mathbf{1}_{\left\{i\in\mathcal{L}^{(n)}\right\}} + \sum_{k=1}^L \mathbb{E}\left[\int_{0} ^T\frac{X_i^{(k)}(t)}{M}\mu^{(k)}dt\right], 
\end{eqnarray}
where we used the fact that $\mathbb{E}[N_p^{(i)}([0,T])]=\lambda^{(i)}T$ because it is a HPP of rate $\lambda^{(i)}$.

$\bullet$ \textbf{Outgoing posts of origin $i$ from Newsfeed $n$:} Each of the posts of any origin entering the Newsfeed of user $n$ (say at time $t$) will replace at random an existing post present on the list. This is the \textit{random eviction} assumption. The evicted post will be of origin $i$ with probability $X_i^{(n)}(t)/M$. We define as above the Bernoulli random variable
\begin{eqnarray}
\label{Zout}
Z_i^{(n)}(t)  & = & \left\{ \begin{tabular}{l l}
$1$ & , with probability $\frac{X_i^{(n)}(t)}{M}$\\
$0$ & , with probability $1-\frac{X_i^{(n)}(t)}{M}$
\end{tabular}
\right..
\end{eqnarray}
This is itself a random process, which depends on the current state of the Newsfeed $n$. The outgoing process of posts of origin $i$ from Newsfeed $n$ results from thinning the total incoming process $N((0,T])$ shown in (\ref{countT}) based on (\ref{Zout}). Using arguments similar to (\ref{expectk}), the expected number of posts of origin $i$ which leave Newsfeed $n$ up to time $T$ is
\begin{eqnarray}
\label{outT}
 & &\hspace{-1.6cm} \mathbb{E}[N_{out,i}((0,T])]= \nonumber\\
 & & =\mathbb{E}\left[\int_{0}^T\frac{X_i^{(n)}(t)}{M}dt\right]\sum_{k=1}^L \left(\lambda^{(k)}+\mu^{(k)}\right).
\end{eqnarray}

$\bullet$ \textbf{Newsfeed conservation law:} The way we modelled the Newsfeed, no post is lost and hence all incoming posts will eventually leave the Newsfeed. The conservation law for posts of origin $i$ up to time $T$ states that
\begin{eqnarray}
\label{conserve}
X_i^{(n)}(0) + N_{in,i}((0,T])  =  N_{out,i}((0,T]) + X_i^{(n)}(T),
\end{eqnarray}
where $X_i^{(n)}(0)$, $X_i^{(n)}(T)$ is the count of origin $i$ posts on the Newsfeed $n$ at time $t=0$ and $t=T$, respectively. 

$\bullet$ \textbf{Taking expectations:} By taking expectation on both sides of the conservation law, we reach the equation 
\begin{eqnarray}
\label{conserveE}
\mathbb{E}[N_{in,i}((0,T])]  & = &  \mathbb{E}[N_{out,i}((0,T])],
\end{eqnarray}
where, $\mathbb{E}[X_i^{(n)}(0)] =  \mathbb{E}[X_i^{(n)}(T)]$ is cancelled out, because we have assumed that at $t=0$ the system is already in steady-state. Let us divide both side in (\ref{conserveE}) by the window size $T$. 
%it appears $\mathbb{E}\left[N_{in,i}((0,T])\right]$ on the left-hand side given in (\ref{inT}), and $\mathbb{E}\left[N_{out,i}((0,T])\right]$ on the right-hand side given in (\ref{outT}).
%After having taken expectation, we divide both sides of the conservation equality (\ref{conserve}) by $T$ and  take the limit as $T\rightarrow\infty$. 
%Then, $\lim_{T\rightarrow\infty}X_i^{(n)}(0) /T =0$ and $\lim_{T\rightarrow\infty}X_i^{(n)}(T) /T =0$. 
At the left-hand side of (\ref{conserveE}) we get using (\ref{inT})%  the conservation equality we get (see (\ref{inT}))
\begin{eqnarray}
\label{inlim}
\lambda^{(i)}\mathbf{1}_{\left\{i\in\mathcal{L}^{(n)}\right\}}+\sum_{k=1}^L \mu^{(k)}\mathbb{E}\left[\frac{1}{T}\int_{0}^T\frac{X_i^{(k)}(t)}{M}dt\right].
\end{eqnarray}
The detailed Markov chain has finite state-space (finite users and size $M$ of Newsfeeds), and we assume that the Follower graph is strongly connected and that $\lambda^{(n)}, \mu^{(n)}>0,\ \forall n\in\mathcal{N}$, hence the chain is ergodic. Since we assume that we observe the chain at its steady-state, then during $(0,T]$ it behaves according to its stationary distribution, and it holds
\begin{eqnarray}
\label{plim}
\mathbb{E}\left[\frac{X_i^{(k)}(t)}{M}\right] & = & p_i^{(k)}, \ \ \ \forall t\in(0,T].
\end{eqnarray}
At the right-hand side of (\ref{expectk}) -- and also in (\ref{inT}), (\ref{outT}) -- we can use Tonelli's theorem to interchange integration and expectation, so that
\begin{eqnarray}
\label{steadystate}
\mathbb{E}\left[\frac{1}{T}\int_{0}^T\frac{X_i^{(k)}(t)}{M}dt\right] = \frac{1}{T}\int_{0}^T \mathbb{E}\left[\frac{X_i^{(k)}(t)}{M}\right]dt=p_i^{(k)}.
\end{eqnarray}
Then (\ref{inlim}) takes the expression at the right-hand side of our balance equation in (\ref{Cpj}), and we can similarly prove the left-hand side of the balance equation with (\ref{outT}) at the steady-state.\\
\qed

\textbf{[C. Lemma for the proof of Theorem~\ref{Main}]} 
\begin{lem}\cite[Chapter 6, Lemma 2.1]{BerPleNN}
\label{Lemma1}
Given a nonnegative matrix $\mathbf{A}\in\mathbb{R}_+^{N\times N}$, its spectral radius is $\rho(\mathbf{A})<1$ if and only if $(\mathbf{I}_N - \mathbf{A})^{-1}$ exists, which can be written as the series
\begin{eqnarray}
\label{series}
(\mathbf{I}_N - \mathbf{A})^{-1} & = & \sum_{n=0}^{\infty} \mathbf{A}^n\geq 0.
\end{eqnarray}
\end{lem}

%====================================
%AUTHORS BIOGRAPHIES

%\begin{IEEEbiography}[{\includegraphics[width=1in,height=1.25in,clip,keepaspectratio]{08_Anastasios.jpg}}]

\begin{IEEEbiography}
{Anastasios Giovanidis (S'03-M'07)} received the Diploma degree in electrical and computer engineering from the National Technical University of Athens, Greece, in 2005, and the Dr.-Ing. degree from the Technical University of Berlin, Germany, in 2010. He has held research positions at the Zuse Institute Berlin, Germany, at the National Institute for Research in Computer Science and Automation (Inria), France, and at T\'el\'ecom ParisTech, France. He is currently a permanent researcher with the French National Center for Scientific Research (CNRS) at LIP6 (Sorbonne Universit\'e, CNRS), Paris, France. His research interest are in the stochastic modelling, data analysis and optimisation of wireless, content delivery and social networks.
\end{IEEEbiography}

\begin{IEEEbiography}%[{\includegraphics[width=1in,height=1.25in,clip,keepaspectratio]{08_Bruno.jpg}}]
{Bruno Baynat}
 received the M.S. degree from the Institut National Polytechnique de Grenoble, France in 1988 and the Ph.D. degree from the University Pierre et Marie Curie, France in 1991. Presently, he is Ma\^itre de Conf\'erence (Associate Professor) at Sorbonne University. His research interests are presently in the development of models for the performance evaluation of communication systems, with applications to wireless and mobile networks, resource allocation and social networks.
\end{IEEEbiography}

\begin{IEEEbiography}%[{\includegraphics[width=1in,height=1.25in,clip,keepaspectratio]{08_Clemence.png}}]
{Cl\'emence Magnien}
 completed her Ph.D. in Computer Science from
Ecole Polytechnique, France in 2003. She currently is a researcher with the
Centre National de la Recherche Scientifique (CNRS) at LIP6,
(Sorbonne Universit\'e, CNRS), Paris, France.
She has been promoted to senior researcher in 2013.
Her research area is the study of large graphs occurring in practice.
She has been particularly interested in recent years in the dynamics
of interactions and has contributed to the design of the link streams
formalism to represent and analyse such dynamics.
\end{IEEEbiography}

\begin{IEEEbiography}%[{\includegraphics[width=1in,height=1.25in,clip,keepaspectratio]{08_Antoine.jpg}}]
{Antoine Vendeville}
graduated in mathematics at Sorbonne Universit\'e, France in 2015, where he received his Master's degree in mathematical modelling in 2018. In 2019 he received a Master's degree in data science from Universit\'e Claude Bernard. He is currently pursuing a PhD in the Computer Science department of University College London, where he is part of the Centre for Doctoral Training in Cybersecurity. His work focuses on mathematical models for opinion dynamics and the development of algorithmic methods to control the echo chamber effect on social networks.
\end{IEEEbiography}

\end{document}